\documentclass[reprint, nofootinbib,amsmath,amssymb,aps]{revtex4-2}
\usepackage{multirow}

\usepackage{graphicx}% Include figure files
\graphicspath{{Figures/}{./}}
\usepackage{dcolumn}% Align table columns on decimal point
\usepackage{bm}% bold math
\usepackage[colorlinks = true,
            linkcolor = black,
            urlcolor  = blue,
            citecolor = blue,
            anchorcolor = black]{hyperref}
\usepackage[mathlines]{}% Enable numbering of text and display math
\usepackage{aas_macros}
\usepackage{xspace}
\usepackage{multirow}
\usepackage{tikz,xcolor}
\usepackage{xspace}
\usepackage{subcaption}
\usepackage{rotating}
\usepackage{orcidlink}

\newcommand{\fermi}{\textit{Fermi}\xspace}
\newcommand{\fermilat}{\fermi-LAT\xspace}
\newcommand{\Fermilat}{\fermi~LAT\xspace}

\begin{document}
%\linenumbers
%\preprint{}
\title{Constraints on the intergalactic magnetic field from \fermilat observations of GRB~221009A}

\author{Lea Burmeister$^{1,2}$}
\author{Paolo Da Vela$^3$\,\orcidlink{0000-0003-0604-4517}}
\email[Contact author: ]{paolo.davela@inaf.it}
\author{Francesco Longo$^{4,5}$\,\orcidlink{0000-0003-2501-2270}}
\author{Guillem Mart\'{i}-Devesa$^{4,5}$\,\orcidlink{0000-0003-0766-6473}}
\email[Contact author: ]{guillem.marti-devesa@ts.infn.it}
\author{Manuel Meyer$^6$\,\orcidlink{0000-0002-0738-7581}}
\email[Contact author: ]{mey@sdu.dk}
\author{Francesco G. Saturni$^{7,8}$}
\author{Antonio Stamerra$^{7}$\;\orcidlink{0000-0002-9430-5264}}
\author{P\'{e}ter Veres$^{9,10}$\;\orcidlink{0000-0002-2149-9846}}

\affiliation{$^1$ Institute for Experimental Physics, University of Hamburg, Luruper Chausee 149, 22761 Hamburg, Germany}
\affiliation{$^2$ Kirchhoff-Institute for Physics, Heidelberg University, 69120 Heidelberg, Germany}
\affiliation{
$^3$INAF – Osservatorio di Astrofisica e Scienza dello Spazio di Bologna, Via Piero Gobetti 93/3, 40129 Bologna, Italy
}
\affiliation{
$^4$ Dipartimento di Fisica, Universit\'a di Trieste, 34127 Trieste, Italy
}
\affiliation{
$^5$ Istituto Nazionale di Fisica Nucleare, Sezione di Trieste, 34127 Trieste, Italy
}
\affiliation{
$^6$CP3-Origins, University of Southern Denmark, Campusvej 55, DK-5230 Odense M, Denmark
}
\affiliation{
$^7$ INAF - Osservatorio Astronomico di Roma, Via Frascati 33, I-00078 Monte Porzio Catone (RM), Italy
}
\affiliation{
$^8$ ASI - Space Science Data Center, Via del Politecnico snc, I-00133 Roma, Italy
}
\affiliation{
$^9$ Department of Space Science, University of Alabama in Huntsville, Huntsville, AL 35899, USA
}
\affiliation{
$^{10}$ Center for Space Plasma and Aeronomic Research (CSPAR), University of Alabama in Huntsville, Huntsville, AL 35899, USA
}

\date{\today}
\begin{abstract}
%An abstract that at most should have 500 words (or $5\%$ of the whole letter)

A cosmological origin of magnetic fields in large scale structures of the Universe would require a non-negligible magnetic field in cosmic voids, which, however, remains undetected.
Gamma-ray emission from gamma-ray bursts (GRBs) offers the opportunity to indirectly probe such an intergalactic magnetic field (IGMF), as $\gamma$ rays interact with cosmic radiation fields, producing electron-positron pairs, and initiate an electromagnetic cascade.
The deflection of the pairs in the IGMF results in a time-delayed signal at GeV energies. 
Using observations with the \textit{Fermi} Large Area Telescope of the GRB\,221009A, we are able to derive the most stringent constraints to date from the nonobservation of the cascade and  rule out magnetic fields $B<2.5\times10^{-17}$\,G at 95$\%$ confidence level for a coherence length larger than 1\,Mpc. 
Our results are comparable to limits obtained from blazar observations but do not suffer from assumptions on the duty cycle of the $\gamma$-ray source or whether inverse-Compton scattering losses dominate over the development of plasma instabilities.

\end{abstract}
%\keywords{}
\maketitle

%\begin{document} 
\section{Introduction \label{sect:intro}}
The origin of the magnetic fields in the large-scale Universe is one of the long-standing problems in cosmology. Magnetic fields with strengths of the order of 1-10 $\mu$G are usually measured in galaxies via observations of Faraday rotation and Zeeman splitting of atomic lines in the radio band. There is general agreement that these magnetic fields can be sourced via amplification mechanisms, such as the $\alpha$-$\omega$ dynamo of weak-seed fields \cite{Kulsrud08}. However, the origin of these weak seeds is largely unknown. Two main hypotheses are usually considered: the \emph{cosmological} and the \emph{astrophysical} origin~\cite{Grasso01,Durrer13}, or a mixture of the two~\cite{Vazza25}.
The main difference between these scenarios is that, in case of cosmological origin, non-negligible magnetic fields are expected in cosmic voids.
Such an intergalactic magnetic field (IGMF) has never been measured and Faraday rotation measurements limit the IGMF strength to be below $10^{-9}$\,G~\cite{Pshirkov16}.

A powerful indirect probe to constrain or measure weak magnetic fields makes use of $\gamma$-ray observations of extragalactic sources (see, e.g., Ref.~\cite{Batista21} for a review). Very-high energy (VHE, $E>100$\,GeV) $\gamma$-ray photons can be absorbed during their propagation by the extragalactic background light (EBL) via the $\gamma$-$\gamma$ pair production process ($\gamma$+$\gamma\rightarrow e^{+}+e^{-}$) \cite{Nikishov62, Gould66}. 
The created pairs inverse Compton (IC) scatter photons from the cosmic microwave background (CMB) up to $\gamma$-ray energies, initiating an electromagnetic cascade.

If an IGMF is present, the pairs are deflected, which induces several observational signatures of the cascade emission, 
such as an additional spectral component and a $\gamma$-ray halo around otherwise pointlike sources.
The absence of these features in $\gamma$-ray observations of active galactic nuclei (AGNs) with their jets closely aligned to the line of sight (so-called blazars) has led to lower limits on the IGMF in the range of 10$^{-15}$--10$^{-17}$\,G e.g., \cite[][]{neronov10,taylor11,dermer11,tavecchio11,ackermann18,Meyer23}.
The results depend on various assumptions such as the intrinsic VHE spectral shape, the AGN duty cycle, as well as the coherence length of the IGMF.
Recently, a lower bound of the order of $\sim$ 10$^{-17}$\,G on IGMF strength was derived, which took into account the variability pattern of the source in the VHE band~\cite{magic22}. 

The discovery of the VHE emission from gamma-ray bursts (GRBs) \cite{magic19_14C,hess18} offers a unique opportunity to constrain the IGMF. 
A non-zero IGMF would cause a time delay of the cascade emissions, producing observable $\gamma$-ray emission even after the GRB afterglow has ceased \cite{Plaga95,Razzaque04,Murase08,Murase09,Takahashi11}. 
This approach has the advantage that it does not require assumptions on the source's duty cycle or variability pattern as these quantities are directly determined for GRBs.

Several attempts have been made to measure the IGMF using this \textit{pair-echo} emission using the first GRB detected in the VHE band, GRB\,190114C~\cite{Wang20,Dzhatdoev20,DaVela23,Vovk23}. 
The authors of Ref. \cite{Wang20} excluded IGMF strengths lower than $\sim$ $10^{-19.5}$ G for a coherence length $\leq$ 1 Mpc, which was however, not confirmed
in other studies \cite{Dzhatdoev20, DaVela23}.
Instead, no limits could be derived. These discrepancies can be explained with the different assumptions made for the intrinsic VHE spectral shape of the GRB and shifting from semi-analytical to cosmic-ray propagation codes for the calculation of the cascade spectral energy distributions (SEDs). Therefore no IGMF constraint can be inferred from this GRB.

Similar studies have been performed using observations of the exceptionally bright  GRB\,221009A  up to 90 days after the GRB trigger~\cite{Huang23,Dzhatdoev24,Vovk23b}.
Importantly, this GRB has been detected at energies larger than 10 TeV %up to TeV energies 
with the Large High Altitude Air Shower Observatory (LHAASO) at high significance~\cite{lhaaso23,lhaaso23b}. In this paper, we report on improved constraints on the IGMF strength from $\gamma$-ray observations with the \fermi Large Area Telescope (LAT) of GRB~221009A during both the afterglow and up to one year after the event.
We use the Monte-Carlo code CRPropa~3 \cite{crpropa3} to simulate the pair-echo SEDs in different time windows and for different IGMF strengths and derive lower bounds on the IGMF employing the full Poissonian likelihood information in the \fermilat energy band.
Throughout the paper, we assume a flat $\Lambda$CDM cosmology with $H_0=70$\,km\,s$^{-1}$\,Mpc$^{-1}$, $\Omega_{\Lambda}$=0.7, and $\Omega_{M}$=0.3.

\section{The VHE intrinsic spectrum of GRB\,221009A}
\label{sect:grb}

The GRB\,221009A was an extremely energetic ($E_{\gamma}^{\rm iso}\sim10^{55}$ erg) event detected with the \fermi gamma-ray burst monitor (GBM) on 2022 October 9 at $T_0=13:16:59.99$ UT \cite{Veres22}. With a redshift $z=0.151$, corresponding to a luminosity distance $D_L \sim 720$ Mpc, its isotropic luminosity is estimated to be $L_{\gamma}^{\rm iso}\simeq10^{53}$ erg s$^{-1}$ \cite{Postigo22}. The event has been subsequently detected with the \Fermilat in the GeV domain \cite{Bissaldi22}, and by the {\it Swift}-BAT and XRT instruments in X-rays \cite{Dichiara22}. Observations started 53 and 55 min after $T_0$, respectively.

At the trigger time $T_0$, the event was in the field of view of LHAASO, which detected more than $6.4\times10^4$ photons within the first 3000 seconds at $E_\gamma > 200$ GeV \cite{lhaaso23}. LHAASO is a hybrid array of VHE detectors, composed of both water Cherenkov photo-multipliers (WCDA) and solid-state scintillators (KM2A) that are sensitive to $\gamma$ rays above a few TeV \cite{Cao19, Xin22}. The intrinsic differential spectra in the range 200\,GeV-7\,TeV, derived within five temporal bins ranging from $T_0 + 5$\,s to $T_0 + 1774$\,s, do not show any evidence of energy cutoff up to 5\,TeV \cite{lhaaso23}. The spectra are compatible with power laws (PLs) with indexes in the range 2.1--2.5.
The KM2A observations reveal $\gamma$ rays with energies up to 13\,TeV~\cite{lhaaso23b}.

In order to predict the correct amount of cascade emission in the GeV energy range, the choice of the VHE intrinsic spectrum is crucial. We follow the approach of Ref.~\cite{DaVela23} and use the physically motivated synchrotron self-Compton (SSC) model fitted to the LHAASO WCDA observations (see Fig.~2 in Ref.~\cite[][]{lhaaso23}).
In each time bin, the SED $E_\gamma^2 F(E_\gamma)$ of the SSC model is well approximated by a log parabola (LP),
\begin{equation}\label{eqn:logpar}
    E_\gamma^2 F(E_\gamma) = S_0 \left(
    \frac{E_\gamma}{E_0}
    \right)^{-\alpha-\eta \log{(E_\gamma/E_0)}}
\end{equation}
with $E_0$ the pivot energy, $S_0$ the SED normalization at $E_\gamma = E_0$, $\alpha$ the PL index and $\eta$ the curvature.
Fixing $E_0 = 1$\,TeV, we derive the LP parameters that approximate the SSC model in each time bin.  
We then compute the time-weighted 
averaged VHE spectrum by taking the geometric mean of the normalization factors and the arithmetic mean of both the spectral index and curvature, weighted by the duration of each time bin.
The average intrinsic SED derived in this way has parameters $S_0 = 6.8 \times 10^{-7}$\,erg\,s$^{-1}$\,cm$^{-2}$, $\alpha =0.38$ and $\eta = 0.10$.
For the subsequent calculations of the expected pair-echo emission, we conservatively incorporate an additional term for an exponential high-energy cutoff $\exp(-E/E_\mathrm{cut})$. We explore the following cases: no cutoff, $E_\mathrm{cut}= 7$~TeV (limit of the WCDA sensitivity), and 13\,TeV (highest energy photon detected with KM2A). For all spectra, we set a maximum energy of $30$\,TeV. % 
We assume that the average emission is well represented by this spectrum over the assumed activity time of $\Delta T_\mathrm{activity}= 1774\,$s, which is the time window over which the SSC modeling has been performed by the LHAASO Collaboration.
\begin{figure*}
    \centering
    \includegraphics[width=.98\linewidth]{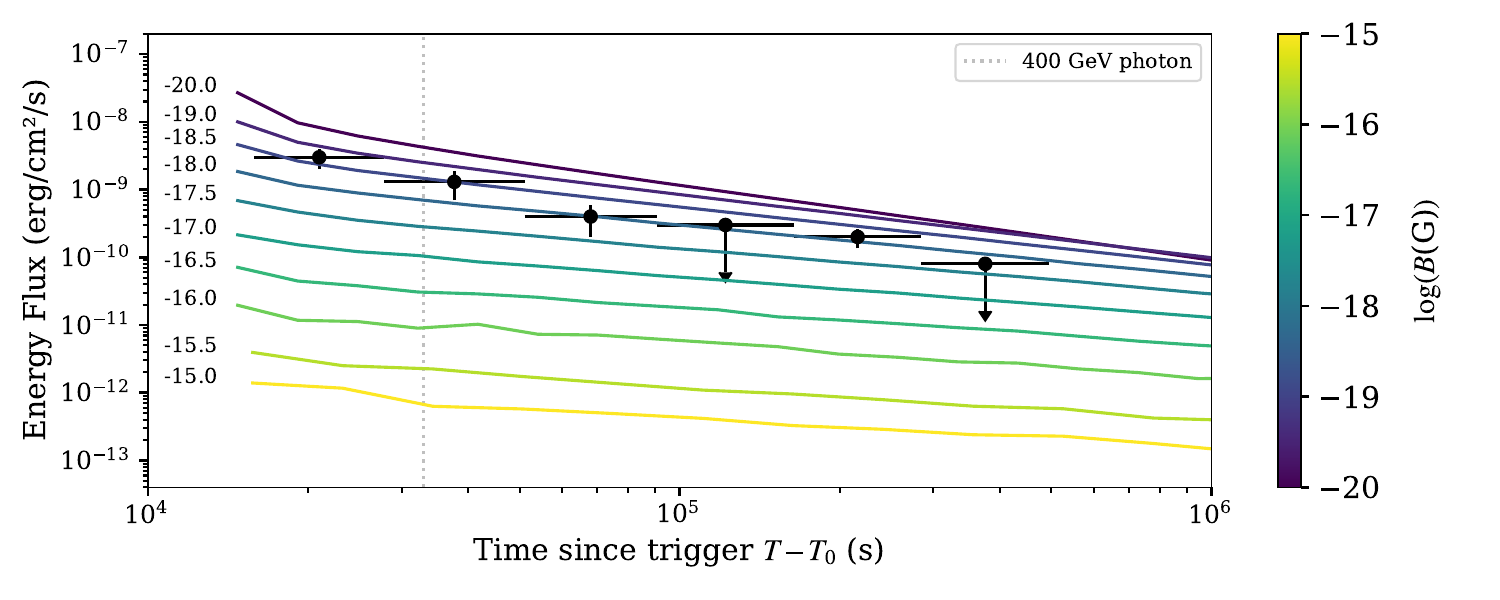}
    \caption{Predicted light curves from the pair-echo emission for $E>100$\,MeV using different magnetic fields, set in context with the LAT fluxes derived in \cite{LATGRB221009A}. Note that the presumed GeV afterglow component lasts until $\sim T_0+2.8\times10^5$ s. The vertical dashed line represents the arrival time of the 400 GeV event not accounted for in standard afterglow emission models. The results were derived with CRPropa simulations employing the log-parabola model of Eq.~\eqref{eqn:logpar} multiplied by an exponential cutoff at 7 TeV as the injected VHE spectrum.}
    \label{fig:light-curve}
\end{figure*}

\begin{figure*}
    \centering
    \includegraphics[width=0.49\linewidth]{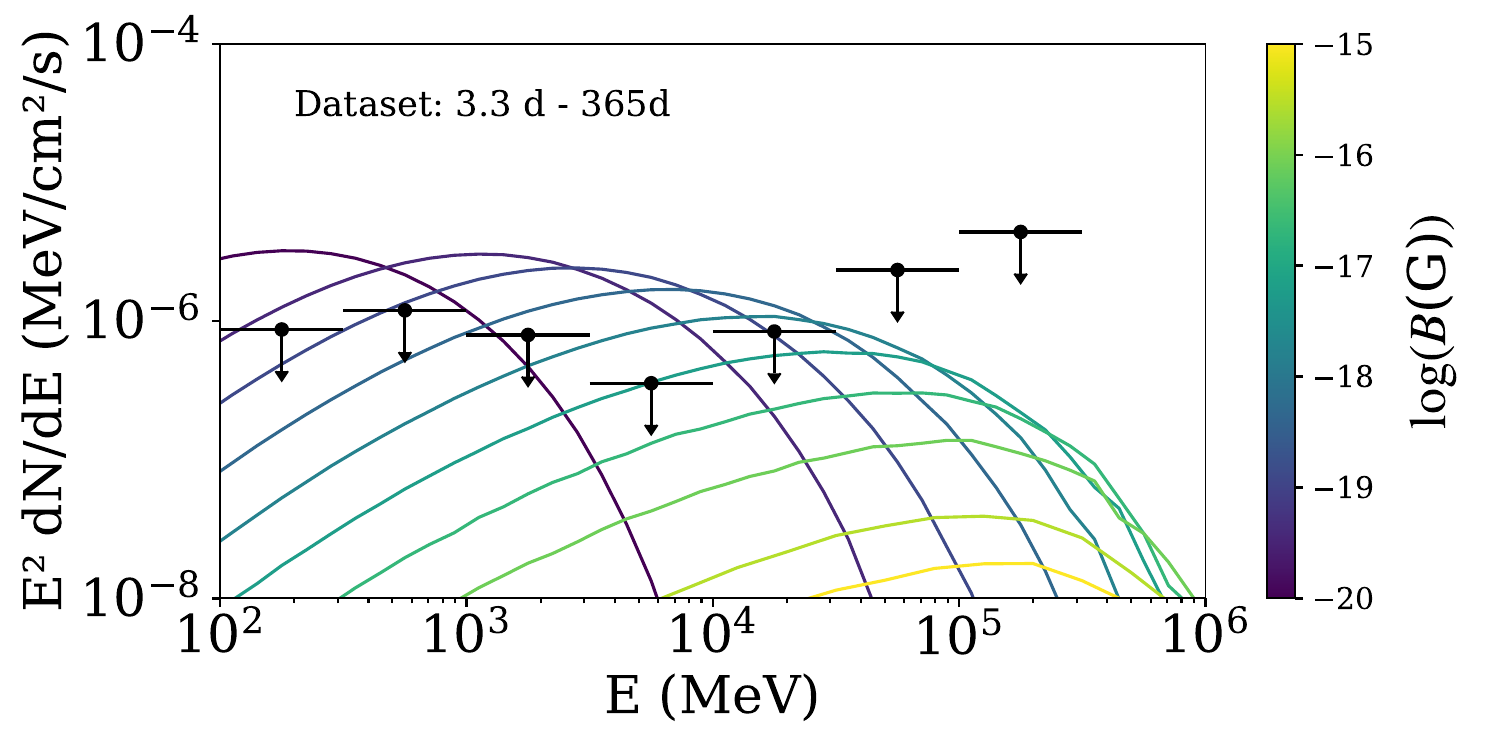}
    \includegraphics[width=0.49\linewidth]{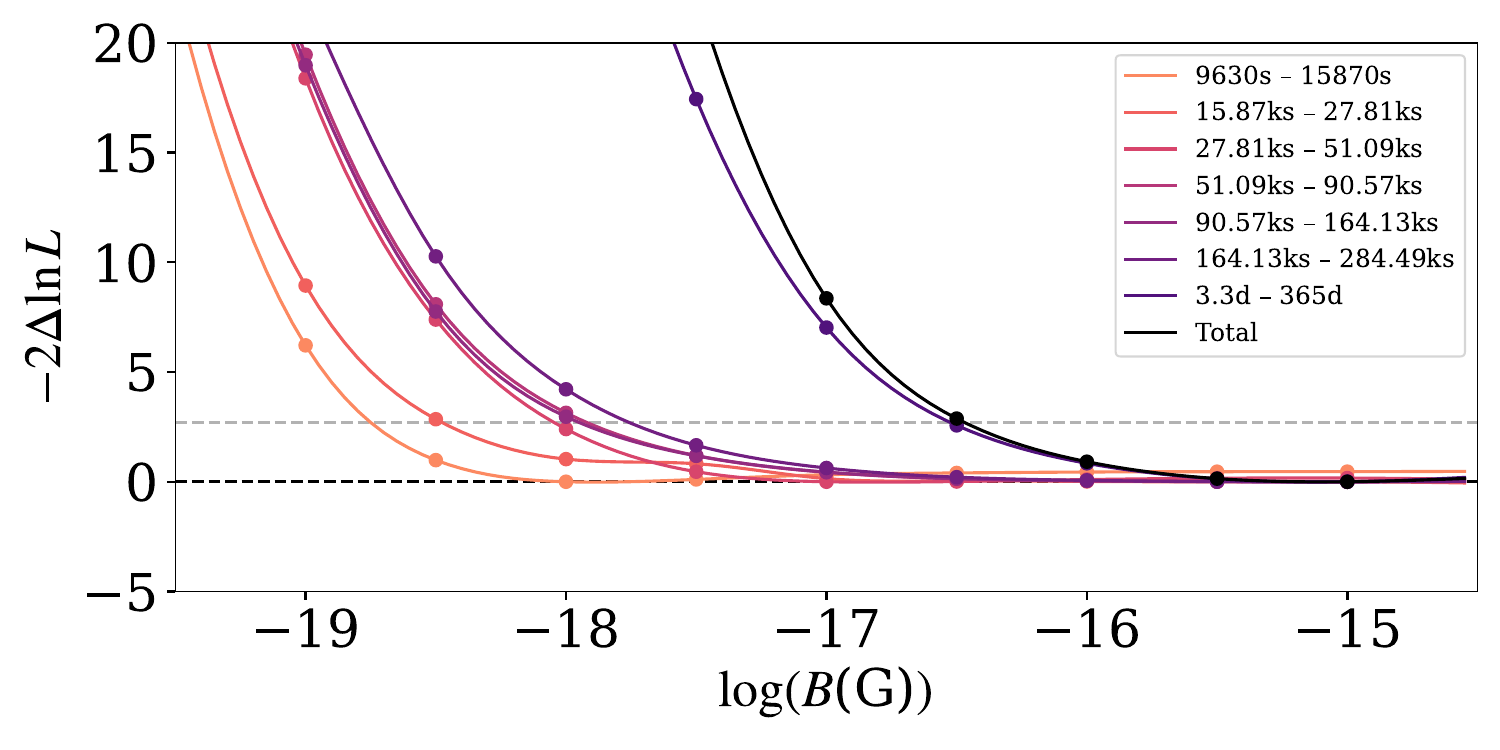}
    \caption{\textit{Left}: simulated SEDs of the pair-echo emission expected between 3.3 and 365 days after $T_0$ with the same injection spectrum as in Fig.~\ref{fig:light-curve} compared with the \fermilat upper limits. \textit{Right}: likelihood profiles sampling the magnetic field for different time intervals. The astrophysical GRB afterglow model assumes an \textit{ad hoc} power law spectrum with $\Gamma=2$. Solid lines represent cubic spline interpolations between the derived likelihood values. The results were derived using the same set of simulations as in Fig.~\ref{fig:light-curve}.}
    \label{fig:cascade_seds}
\end{figure*}

\section{Simulation of the pair-echo emission}
\label{sect:crpropa}

To produce the pair-echo SEDs in different time windows we employ the Monte-Carlo code CRPropa~3 (version 3.2),
which simulates the development of the pair-echo emission given a primary VHE photon spectrum injected in the intergalactic medium \cite{crpropa3}.
The source is located at the center of a sphere with radius $D$ equal to the comoving distance to the source\footnote{A redshift $z=0.151$ is sufficiently nearby for a reliable EBL implementation within CRPropa3, see Figure B.1 in \citep{Kalashev23}.}. We assume that the source emits VHE photons in the range 0.1-30 TeV according to Eq.~\eqref{eqn:logpar}, and re-weight the SED accordingly when a cutoff is adopted. These photons are injected considering an aperture cone of 1.6$^{\circ}$, as inferred from the break in the LHAASO light curve~\cite{lhaaso23}. The choice of the cone aperture is based on the GRB emission model adopted by the LHAASO collaboration. We note that a larger opening angle could affect the cascade flux only for large time delays. However, for the time delays considered in this work, this parameter does not significantly impact the results.

Examples of the derived pair-echo light curves and SEDs are shown in Figs.~\ref{fig:light-curve}~and~\ref{fig:cascade_seds}.
In Fig.~\ref{fig:light-curve}, we show the cascade flux %integrated over 
for $E>100$~MeV as a function of time for different IGMF strengths for a cutoff in the intrinsic spectrum of 7\,TeV.
The left panel of Fig.~\ref{fig:cascade_seds} shows the SED averaged between 3.3 days and 365 days. 
The lower value of 3.3\,days corresponds to the last detection from the GRB as reported by \Fermilat \citep{LATGRB221009A}. 
As expected, both figures show that the cascade flux drops with increasing magnetic field, since the time delay of the cascade photons increases with increasing IGMF strength. 
Furthermore, the peak of the cascade flux moves towards higher energies with increasing IGMF. This is caused by the energy dependence of the deflection angle of the pairs: for the same magnetic field $B$, the deflection angle is proportional to $E^{-2}$ for large correlation lengths \citep{Neronov09}.

This is due to the longer time delay of low energy photons as low energy pairs have a larger Larmor radius for higher IGMF values.
We provide further details on the CRPropa simulations in Appendix \ref{app:CRPROPA}.

\section{\textit{Fermi}-LAT limits on the pair echo emission}

The \fermi~LAT is a pair-conversion telescope that detects $\gamma$ rays from 20\,MeV to beyond 300\,GeV \citep{Fermi}. We select events between 100 MeV and 1 TeV, within a $10^{\circ}\times10^{\circ}$ region centered on GRB 221009A, with a zenith angle cut at $90^{\circ}$. We divide the dataset in several time bins following the main analysis of the \fermilat Collaboration \citep{LATGRB221009A}, expanding it up to one year after $T_0$ (Appendix \ref{app:LAT}).

The likelihood of observing the data $\mathcal{D}_{ij}$ in the $i$th time bin and $j$th energy bin follows a Poisson likelihood $\mathcal{L}(B, \boldsymbol{\theta}_i | \mathcal{D}_{ij})$. This assumes an IGMF strength $B$ and spectral parameters $\boldsymbol{\theta}$ for both the background sources and potential GRB afterglow emission. Background sources can be nearby point sources or diffuse backgrounds. The log-likelihood ratio test can then be written as

\begin{equation}
\label{eq:likelihood}
    \lambda(B) = -2 \sum\limits_{i,j}\ln\left(\frac{\mathcal{L}(B, \hat{\hat{\boldsymbol{\theta}}}_i | \mathcal{D}_{ij})}{\mathcal{L}(\hat{B}, \hat{\boldsymbol{\theta}}_i | \mathcal{D}_i)}\right),
\end{equation}
where $\hat{\hat{\boldsymbol{\theta}}}_i$ denotes the spectral parameters maximizing $\mathcal{L}_i$ for a fixed IGMF field strength in the $i$th time bin, whereas $\hat{\boldsymbol{\theta}}_i$ and $\hat{B}$ are the parameters that maximize $\mathcal{L}_i$ unconditionally. Note that, for time bins earlier than $T_0 + 3.3\,$~days, we additionally included a point source with a power law spectrum with index $\Gamma=2$ and free normalization to account for the GRB afterglow emission (see Appendixes~\ref{app:LAT} and~\ref{app:SSCmodel}).

We show the results of the log-likelihood ratio test in the right panel of Fig.~\ref{fig:cascade_seds}. 
No IMGF strength is preferred. 
Instead, we are able to rule out IGMF strengths at 95\,\% confidence for which $\lambda > 2.71$. 
Combining all time and energy bins, this excludes $B<2.5\times10^{-17}$~G.
The most stringent constraint comes from the latest time bin (up to $T_0 + 1$ yr), which rules out $B<2.2\times10^{-17}$~G.
Constraints from the individual time bins are reported in Appendix~\ref{app:LAT}.

\section{Discussion and conclusions \label{sect:discussion}}

Our results depend on the chosen high-energy cutoff of the injection spectrum of VHE $\gamma$~rays as well as the assumed GRB afterglow emission in the early time bins. 
Increasing the cutoff energy to 13\,TeV, corresponding to the highest energy reported with the LHAASO KM2A detectors rules out $B < 3.5\times 10^{-17}$\,G. 
This is expected, as photons injected at even higher energies will increase the amount of pair-echo photons in the \fermilat energy band. 
If no cutoff is applied to the log-parabola spectrum the constraints further improve and rule out an IGMF with $B < 5\times 10^{-17}$\,G.
These results assume that an afterglow component can account for the GeV signal, fitting a power law with $\Gamma=2$ to the flux observed by the LAT in each time bin. To estimate the impact of assuming such an \textit{ad hoc} component, we alternatively test a scenario where we replace the simple power law model with a physically motivated SSC model. This model is set to reproduce the LHAASO data, and only employed to predict the GeV flux in the $i$th time bin (Appendix~\ref{app:SSCmodel}).
The limits weaken marginally, and an IGMF with $B < 1.4\times10^{-17}\,$G is ruled out for a high-energy cutoff at 7\,TeV. 
Finally, we also considered the scenario of no astrophysical afterglow.
In this case, a preference for the cascade is found in the early time bins in order to explain the \Fermilat detection. 
However, different time bins prefer different values of the IGMF, which are all incompatible with the constraints from the latest time bin extending up to $T_0 + 1\,$yr. 
Therefore, we do not regard this as evidence for an IGMF. 

\begin{figure}
    \centering
    \includegraphics[width=0.99\linewidth]{GRB221009_constraints.pdf}
    \caption{The IGMF parameter space ($B$, $\ell_B$). Theoretically preferred regions by electroweak (EWPT) or QCD phase transitions (QCDPT), or by galactic winds and induced by cosmic-ray (CR) streaming are shown in blue~\cite{Durrer13}. Previous constraints using LAT observations of GRB~221009A, as well as constraints from $\gamma$-ray observations of blazars are shown as gray lines and gray shaded regions with arrows. Our constraints at the 95\,\% confidence level are shown in orange.}
    \label{fig:constraints}
\end{figure}

In principle, the choice of the EBL model could affect the amount of cascade flux. In this work, we used the EBL model by \cite{Franceschini08}. However, as we show in Appendix~\ref{app:CRPROPA}, the cascade flux level does not vary significantly across different EBL models, so the IGMF limit is not sensitive to the particular choice of model.

Interestingly, the LAT detected a 400 GeV photon at $T_0+33$ ks, whose detection is inconsistent with a synchrotron origin \citep{LATGRB221009A}. While it has been suggested that it could be attributed to the cascade emission (see e.g. \citep{400GeV}), our results indicate otherwise. The 400 GeV flux cascade at 33 ks is predicted to be extremely low, with probabilities of detecting a single photon above 100 GeV of less than $3\sigma$.

Our results are significantly more constraining (by more than an order of magnitude) than previous analyses also using \fermilat observations of 
GRB~221009A~\cite{Huang23,Dzhatdoev24,Vovk23b}, see Fig.~\ref{fig:constraints}. 
The most significant factor concerns the time window considered for searching the pair echo. Previously, the cascade signal was searched for up to 20\,days~\cite{Vovk23b}, 90\,days~\cite{Dzhatdoev24} or eight\,months~\cite{Huang23} after $T_0$. 
Furthermore, in our analysis we make full use of both spectral and temporal information given the chosen time binning.

Given the exceptionally high flux in the VHE band, it is plausible that \fermilat observations could still be sensitive to the pair-echo emission at later times of more than one year.  We assess this possibility by extending our simulations and extrapolating the LAT exposure up to five\,years. 
Given the energy-dependent dilution in time of the signal, we find that the average flux expected within the 3--15\,GeV band will tighten the constraints if several years of exposure are added, despite not reaching an additional order of magnitude (see Appendix~\ref{app:LAT}). 

The lower bound on the IGMF strength derived in this work holds as long as the correlation length $\ell_B$ of the IGMF is larger than the cooling distance of the electron-positron pairs for IC scattering, $l_{IC}\simeq 0.7\left(E_\gamma/ ~\rm{TeV}\right)^{-1}$ for primary $\gamma$~rays with energy $E_\gamma$.
Considering $E_\gamma\sim 10\,$TeV, we find $\ell_B<l_\mathrm{IC}$ for $\ell_B \sim 0.1\,$Mpc, under the assumption that the electron and positron have both half the energy of the $\gamma$~ray.
For lower values of the coherence length, the lower bound on the IGMF strength should scale as $\ell_B^{-1/2}$ for a Kolmogorov turbulence spectrum~\cite{Neronov09}, as electron-positron stochastic deflections can be described as diffusion in angle
(see Ref.~\cite{Caprini2015} for the scaling with different turbulence spectra).
Our result represents the best constraint so far on the IGMF employing the pair-echo technique, comparable to those obtained by studying blazars~\cite{magic22}. This opens the door for IGMF studies by means of TeV-detected GRBs to be as competitive as those done on persistent extragalactic VHE sources, while not suffering from assumptions on the duty cycles~\cite{Meyer23}.
Moreover, our results are not affected by the possible development of plasma instabilities in the electron-positron beam in voids \citep{Broderick12}. 
It has been argued that such instabilities could lead to an energy loss of the pairs on timescales shorter than the IC cooling time. However, the time required for such instabilities to grow is 1 order of magnitude larger than the time activity of the GRB 221009A, even with its large TeV luminosity \cite{Broderick12, lhaaso23}.\\

\begin{acknowledgments}

The \textit{Fermi}-LAT Collaboration acknowledges support for LAT development, operation and data analysis from NASA and DOE (United States), CEA/Irfu and IN2P3/CNRS (France), ASI and INFN (Italy), MEXT, KEK, and JAXA (Japan), and the K.A.~Wallenberg Foundation, the Swedish Research Council and the National Space Board (Sweden). Science analysis support in the operations phase from INAF (Italy) and CNES (France) is also gratefully acknowledged. This work performed in part under DOE Contract DE-AC02-76SF00515. P.V. acknowledges support from NASA grant NNM11AA01A.
M.M. acknowledges support from the European Research Council (ERC) under the European Union’s Horizon 2020 research and innovation program Grant agreement No. 948689 (AxionDM). L.B. acknowledges support from the Deutsche Forschungsgemeinschaft (DFG, German Research Foundation) under Germany’s Excellence Strategy—EXC 2121 ``Quantum Universe''. P.D.V. acknowledges “funding by the European Union – NextGenerationEU” RFF M4C2 project IR0000012 CTA+. We acknowledge financial support from INAF through the “Ricerca Fondamentale 2024” program (mini-grant titled ‘Measurement of the intergalactic magnetic field with gamma-ray sources’ (PI A.S.).

\end{acknowledgments}

\bibliography{GRB-IGMF}

@ARTICLE{Sari+01ic,
   author = {{Sari}, R. and {Esin}, A.~A.},
    title = "{On the Synchrotron Self-Compton Emission from Relativistic Shocks and Its Implications for Gamma-Ray Burst Afterglows}",
  journal = {\apj},
   eprint = {arXiv:astro-ph/0005253},
     year = 2001,
    month = feb,
   volume = 548,
    pages = {787-799},
      doi = {10.1086/319003},
   adsurl = {http://adsabs.harvard.edu/abs/2001ApJ...548..787S}
}

@ARTICLE{panaitescu00,
   author = {{Panaitescu}, A. and {Kumar}, P.},
    title = "{Analytic Light Curves of Gamma-Ray Burst Afterglows: Homogeneous versus Wind External Media}",
  journal = {\apj},
   eprint = {arXiv:astro-ph/0003246},
     year = 2000,
    month = nov,
   volume = 543,
    pages = {66-76},
      doi = {10.1086/317090},
   adsurl = {http://adsabs.harvard.edu/abs/2000ApJ...543...66P}
}

@ARTICLE{chiangdermer99,
   author = {{Chiang}, J. and {Dermer}, C.~D.},
    title = "{Synchrotron and Synchrotron Self-Compton Emission and the Blast-Wave Model of Gamma-Ray Bursts}",
  journal = {\apj},
   eprint = {arXiv:astro-ph/9803339},
     year = 1999,
    month = feb,
   volume = 512,
    pages = {699-710},
      doi = {10.1086/306789},
   adsurl = {http://adsabs.harvard.edu/abs/1999ApJ...512..699C}
}

@ARTICLE{Grasso01,
       author = {{Grasso}, D. and {Rubinstein}, H.~R.},
        title = "{Magnetic fields in the early Universe}",
      journal = {Phys. Rep.},
         year = 2001,
        month = jul,
       volume = {348},
       number = {3},
        pages = {163-266},
          doi = {10.1016/S0370-1573(00)00110-1},
archivePrefix = {arXiv},
       eprint = {astro-ph/0009061},
 primaryClass = {astro-ph},
       adsurl = {https://ui.adsabs.harvard.edu/abs/2001PhR...348..163G}
}

@ARTICLE{Vazza25,
       author = {{Vazza}, F. and {Gheller}, C. and {Zanetti}, F and {Tsizh}, M. and {Carretti}, E. and {Mtchedlidze}, S. and {Brueggen}, M.},
        title = "{The evolution of cosmic ray electrons in the cosmic web: seeding by AGN, star formation and shocks}",
      journal = {arXiv e-prints},
         year = 2025,
        month = jan,
          eid = {arXiv:2501.19041},
        pages = {arXiv:2501.19041},
          doi = {10.48550/arXiv.2501.19041},
archivePrefix = {arXiv},
       eprint = {2501.19041},
 primaryClass = {astro-ph.HE},
       adsurl = {https://ui.adsabs.harvard.edu/abs/2025arXiv250119041V}
}

@ARTICLE{Pshirkov16,
       author = {{Pshirkov}, M.~S. and {Tinyakov}, P.~G. and {Urban}, F.~R.},
        title = "{New Limits on Extragalactic Magnetic Fields from Rotation Measures}",
      journal = {\prl},
         year = 2016,
        month = may,
       volume = {116},
       number = {19},
          eid = {191302},
        pages = {191302},
          doi = {10.1103/PhysRevLett.116.191302},
archivePrefix = {arXiv},
       eprint = {1504.06546},
 primaryClass = {astro-ph.CO},
       adsurl = {https://ui.adsabs.harvard.edu/abs/2016PhRvL.116s1302P}
}

@ARTICLE{Batista21,
       author = {{Alves Batista}, Rafael and {Saveliev}, Andrey},
        title = "{The Gamma-Ray Window to Intergalactic Magnetism}",
      journal = {Universe},
         year = 2021,
        month = jul,
       volume = {7},
       number = {7},
          eid = {223},
        pages = {223},
          doi = {10.3390/universe7070223},
archivePrefix = {arXiv},
       eprint = {2105.12020},
 primaryClass = {astro-ph.HE},
       adsurl = {https://ui.adsabs.harvard.edu/abs/2021Univ....7..223A}
}

@ARTICLE{Durrer13,
       author = {{Durrer}, Ruth and {Neronov}, Andrii},
        title = "{Cosmological magnetic fields: their generation, evolution and observation}",
      journal = {Astron. \& Astrophys. Rev.},
         year = 2013,
        month = jun,
       volume = {21},
          eid = {62},
        pages = {62},
          doi = {10.1007/s00159-013-0062-7},
archivePrefix = {arXiv},
       eprint = {1303.7121},
 primaryClass = {astro-ph.CO},
       adsurl = {https://ui.adsabs.harvard.edu/abs/2013A&ARv..21...62D}
}

@ARTICLE{Nikishov62,
       author = {{Nikishov}, A. I.},
        title = "{ABSORPTION OF HIGH-ENERGY PHOTONS IN THE UNIVERSE}",
      journal = {Sov. Phys.--JETP},
         year = 1962,
        month = feb,
       volume = {14},
       number = {2},
        pages = {393-394}
}

@ARTICLE{Gould66,
       author = {{Gould}, Robert J. and {Schr{\'e}der}, Gerald},
        title = "{Opacity of the Universe to High-Energy Photons}",
      journal = {\prl},
         year = 1966,
        month = feb,
       volume = {16},
       number = {6},
        pages = {252-254},
          doi = {10.1103/PhysRevLett.16.252},
       adsurl = {https://ui.adsabs.harvard.edu/abs/1966PhRvL..16..252G}
}

@ARTICLE{Neronov09,
       author = {{Neronov}, A. and {Semikoz}, D.~V.},
        title = "{Sensitivity of {\ensuremath{\gamma}}-ray telescopes for detection of magnetic fields in the intergalactic medium}",
      journal = {Phys. Rev. D},
         year = 2009,
        month = dec,
       volume = {80},
       number = {12},
          eid = {123012},
        pages = {123012},
          doi = {10.1103/PhysRevD.80.123012},
archivePrefix = {arXiv},
       eprint = {0910.1920},
 primaryClass = {astro-ph.CO},
       adsurl = {https://ui.adsabs.harvard.edu/abs/2009PhRvD..80l3012N}
}

@ARTICLE{neronov10,
       author = {{Neronov}, Andrii and {Vovk}, Ievgen},
        title = "{Evidence for Strong Extragalactic Magnetic Fields from Fermi Observations of TeV Blazars}",
      journal = {Science},
         year = 2010,
        month = apr,
       volume = {328},
       number = {5974},
        pages = {73},
          doi = {10.1126/science.1184192},
archivePrefix = {arXiv},
       eprint = {1006.3504},
 primaryClass = {astro-ph.HE},
       adsurl = {https://ui.adsabs.harvard.edu/abs/2010Sci...328...73N}
}

@ARTICLE{taylor11,
       author = {{Taylor}, A.~M. and {Vovk}, I. and {Neronov}, A.},
        title = "{Extragalactic magnetic fields constraints from simultaneous GeV-TeV observations of blazars}",
      journal = {\aap},
         year = 2011,
        month = may,
       volume = {529},
          eid = {A144},
        pages = {A144},
          doi = {10.1051/0004-6361/201116441},
archivePrefix = {arXiv},
       eprint = {1101.0932},
 primaryClass = {astro-ph.HE},
       adsurl = {https://ui.adsabs.harvard.edu/abs/2011A&A...529A.144T}
}

@ARTICLE{dermer11,
       author = {{Dermer}, Charles D. and {Cavadini}, Massimo and {Razzaque}, Soebur and {Finke}, Justin D. and {Chiang}, James and {Lott}, Benoit},
        title = "{Time Delay of Cascade Radiation for TeV Blazars and the Measurement of the Intergalactic Magnetic Field}",
      journal = {\apjl},
         year = 2011,
        month = jun,
       volume = {733},
       number = {2},
        pages = {L21},
          doi = {10.1088/2041-8205/733/2/L21},
archivePrefix = {arXiv},
       eprint = {1011.6660},
 primaryClass = {astro-ph.HE},
       adsurl = {https://ui.adsabs.harvard.edu/abs/2011ApJ...733L..21D}
}

@ARTICLE{tavecchio11,
       author = {{Tavecchio}, F. and {Ghisellini}, G. and {Bonnoli}, G. and {Foschini}, L.},
        title = "{Extreme TeV blazars and the intergalactic magnetic field}",
      journal = {\mnras},
         year = 2011,
        month = jul,
       volume = {414},
       number = {4},
        pages = {3566-3576},
          doi = {10.1111/j.1365-2966.2011.18657.x},
archivePrefix = {arXiv},
       eprint = {1009.1048},
 primaryClass = {astro-ph.HE},
       adsurl = {https://ui.adsabs.harvard.edu/abs/2011MNRAS.414.3566T}
}

@ARTICLE{ackermann18,
       author = {{Ackermann}, M. and {Ajello}, M. and {Baldini}, L. and {Ballet}, J. and {Barbiellini}, G. and {Bastieri}, D. and {Bellazzini}, R. and {Bissaldi}, E. and {Blandford}, R.~D. and {Bloom}, E.~D. and {Bonino}, R. and {Bottacini}, E. and {Brandt}, T.~J. and {Bregeon}, J. and {Bruel}, P. and {Buehler}, R. and {Cameron}, R.~A. and {Caputo}, R. and {Caraveo}, P.~A. and {Castro}, D. and {Cavazzuti}, E. and {Charles}, E. and {Cheung}, C.~C. and {Chiaro}, G. and {Ciprini}, S. and {Cohen-Tanugi}, J. and {Costantin}, D. and {Cutini}, S. and {D'Ammando}, F. and {de Palma}, F. and {Desai}, A. and {Di Lalla}, N. and {Di Mauro}, M. and {Di Venere}, L. and {Favuzzi}, C. and {Finke}, J. and {Franckowiak}, A. and {Fukazawa}, Y. and {Funk}, S. and {Fusco}, P. and {Gargano}, F. and {Gasparrini}, D. and {Giglietto}, N. and {Giordano}, F. and {Giroletti}, M. and {Green}, D. and {Grenier}, I.~A. and {Guillemot}, L. and {Guiriec}, S. and {Hays}, E. and {Hewitt}, J.~W. and {Horan}, D. and {J{\'o}hannesson}, G. and {Kensei}, S. and {Kuss}, M. and {Larsson}, S. and {Latronico}, L. and {Lemoine-Goumard}, M. and {Li}, J. and {Longo}, F. and {Loparco}, F. and {Lovellette}, M.~N. and {Lubrano}, P. and {Magill}, J.~D. and {Maldera}, S. and {Manfreda}, A. and {Mazziotta}, M.~N. and {McEnery}, J.~E. and {Meyer}, M. and {Mizuno}, T. and {Monzani}, M.~E. and {Morselli}, A. and {Moskalenko}, I.~V. and {Negro}, M. and {Nuss}, E. and {Omodei}, N. and {Orienti}, M. and {Orlando}, E. and {Ormes}, J.~F. and {Palatiello}, M. and {Paliya}, V.~S. and {Paneque}, D. and {Perkins}, J.~S. and {Persic}, M. and {Pesce-Rollins}, M. and {Piron}, F. and {Porter}, T.~A. and {Principe}, G. and {Rain{\`o}}, S. and {Rando}, R. and {Rani}, B. and {Razzaque}, S. and {Reimer}, A. and {Reimer}, O. and {Reposeur}, T. and {Sgr{\`o}}, C. and {Siskind}, E.~J. and {Spandre}, G. and {Spinelli}, P. and {Suson}, D.~J. and {Tajima}, H. and {Thayer}, J.~B. and {Tibaldo}, L. and {Torres}, D.~F. and {Tosti}, G. and {Valverde}, J. and {Venters}, T.~M. and {Vogel}, M. and {Wood}, K. and {Wood}, M. and {Zaharijas}, G. and {Fermi-LAT Collaboration} and {Biteau}, J.},
        title = "{The Search for Spatial Extension in High-latitude Sources Detected by the Fermi Large Area Telescope}",
      journal = {\apjs},
         year = 2018,
        month = aug,
       volume = {237},
       number = {2},
          eid = {32},
        pages = {32},
          doi = {10.3847/1538-4365/aacdf7},
archivePrefix = {arXiv},
       eprint = {1804.08035},
 primaryClass = {astro-ph.HE},
       adsurl = {https://ui.adsabs.harvard.edu/abs/2018ApJS..237...32A}
}

@ARTICLE{Meyer23,
       author = {{Aharonian}, F. and {Aschersleben}, J. and {Backes}, M. and {Martins}, V. Barbosa and {Batzofin}, R. and {Becherini}, Y. and {Berge}, D. and {Bi}, B. and {Bouyahiaoui}, M. and {Breuhaus}, M. and {Brose}, R. and {Brun}, F. and {Bruno}, B. and {Bulik}, T. and {Burger-Scheidlin}, C. and {Bylund}, T. and {Caroff}, S. and {Casanova}, S. and {Celic}, J. and {Cerruti}, M. and {Chand}, T. and {Chandra}, S. and {Chen}, A. and {Chibueze}, J. and {Chibueze}, O. and {Cotter}, G. and {de Bony}, M. and {Egberts}, K. and {Ernenwein}, J. -P. and {Fichet de Clairfontaine}, G. and {Filipovic}, M. and {Fontaine}, G. and {F{\"u}ssling}, M. and {Funk}, S. and {Gabici}, S. and {Ghafourizadeh}, S. and {Giavitto}, G. and {Glawion}, D. and {Glicenstein}, J.~F. and {Goswami}, P. and {Grondin}, M. -H. and {Haerer}, L. and {Holch}, T.~L. and {Holler}, M. and {Horns}, D. and {Jamrozy}, M. and {Jankowsky}, F. and {Joshi}, V. and {Jung-Richardt}, I. and {Kasai}, E. and {Katarzy{\'n}ski}, K. and {Khatoon}, R. and {Kh{\'e}lifi}, B. and {Klu{\'z}niak}, W. and {Komin}, Nu. and {Kosack}, K. and {Kostunin}, D. and {Lang}, R.~G. and {Le Stum}, S. and {Leitl}, F. and {Lemi{\`e}re}, A. and {Lenain}, J. -P. and {Leuschner}, F. and {Lohse}, T. and {Luashvili}, A. and {Lypova}, I. and {Mackey}, J. and {Malyshev}, D. and {Malyshev}, D. and {Marandon}, V. and {Marchegiani}, P. and {Marcowith}, A. and {Mart{\'\i}-Devesa}, G. and {Marx}, R. and {Meyer}, M. and {Mitchell}, A. and {Moderski}, R. and {Mohrmann}, L. and {Montanari}, A. and {Moulin}, E. and {Muller}, J. and {Murach}, T. and {Nakashima}, K. and {Niemiec}, J. and {Ohm}, S. and {Olivera-Nieto}, L. and {de Ona Wilhelmi}, E. and {Panny}, S. and {Panter}, M. and {Parsons}, R.~D. and {Peron}, G. and {Prokhorov}, D.~A. and {Prokoph}, H. and {P{\"u}hlhofer}, G. and {Punch}, M. and {Quirrenbach}, A. and {Reichherzer}, P. and {Reimer}, A. and {Reimer}, O. and {Reville}, B. and {Rieger}, F. and {Rowell}, G. and {Rudak}, B. and {Ruiz-Velasco}, E. and {Sahakian}, V. and {Sanchez}, D.~A. and {Sasaki}, M. and {Sch{\"u}ssler}, F. and {Schutte}, H.~M. and {Schwanke}, U. and {Shapopi}, J.~N.~S. and {Sol}, H. and {Spencer}, S. and {Steinmassl}, S. and {Suzuki}, H. and {Takahashi}, T. and {Tanaka}, T. and {Taylor}, A.~M. and {Terrier}, R. and {Thorpe-Morgan}, C. and {Tsirou}, M. and {Tsuji}, N. and {Uchiyama}, Y. and {van Eldik}, C. and {Veh}, J. and {Venter}, C. and {Wagner}, S.~J. and {White}, R. and {Wierzcholska}, A. and {Wong}, Yu Wun and {Zacharias}, M. and {Zargaryan}, D. and {Zdziarski}, A.~A. and {Zouari}, S. and {{\.Z}ywucka}, N. and {Meyer}, M. and {Fermi-LAT Collaboration}},
        title = "{Constraints on the Intergalactic Magnetic Field Using Fermi-LAT and H.E.S.S. Blazar Observations}",
      journal = {\apjl},
         year = 2023,
        month = jun,
       volume = {950},
       number = {2},
          eid = {L16},
        pages = {L16},
          doi = {10.3847/2041-8213/acd777},
archivePrefix = {arXiv},
       eprint = {2306.05132},
 primaryClass = {astro-ph.HE},
       adsurl = {https://ui.adsabs.harvard.edu/abs/2023ApJ...950L..16A}
}

@ARTICLE{magic22,
       author = {{Acciari}, V.~A. and {Agudo}, I. and {Aniello}, T. and {Ansoldi}, S. and {Antonelli}, L.~A. and {Arbet Engels}, A. and {Artero}, M. and {Asano}, K. and {Baack}, D. and {Babi{\'c}}, A. and {Baquero}, A. and {Barres de Almeida}, U. and {Barrio}, J.~A. and {Batkovi{\'c}}, I. and {Becerra Gonz{\'a}lez}, J. and {Bednarek}, W. and {Bernardini}, E. and {Bernardos}, M. and {Berti}, A. and {Besenrieder}, J. and {Bhattacharyya}, W. and {Bigongiari}, C. and {Biland}, A. and {Blanch}, O. and {B{\"o}kenkamp}, H. and {Bonnoli}, G. and {Bo{\v{s}}njak}, {\v{Z}}. and {Burelli}, I. and {Busetto}, G. and {Carosi}, R. and {Ceribella}, G. and {Cerruti}, M. and {Chai}, Y. and {Chilingarian}, A. and {Cikota}, S. and {Colombo}, E. and {Contreras}, J.~L. and {Cortina}, J. and {Covino}, S. and {D'Amico}, G. and {D'Elia}, V. and {da Vela}, P. and {Dazzi}, F. and {de Angelis}, A. and {de Lotto}, B. and {Del Popolo}, A. and {Delfino}, M. and {Delgado}, J. and {Delgado Mendez}, C. and {Depaoli}, D. and {di Pierro}, F. and {di Venere}, L. and {Do Souto Espi{\~n}eira}, E. and {Dominis Prester}, D. and {Donini}, A. and {Dorner}, D. and {Doro}, M. and {Elsaesser}, D. and {Fallah Ramazani}, V. and {Fari{\~n}a}, L. and {Fattorini}, A. and {Font}, L. and {Fruck}, C. and {Fukami}, S. and {Fukazawa}, Y. and {Garc{\'\i}a L{\'o}pez}, R.~J. and {Garczarczyk}, M. and {Gasparyan}, S. and {Gaug}, M. and {Giglietto}, N. and {Giordano}, F. and {Gliwny}, P. and {Godinovi{\'c}}, N. and {Green}, J.~G. and {Green}, D. and {Hadasch}, D. and {Hahn}, A. and {Hassan}, T. and {Heckmann}, L. and {Herrera}, J. and {Hrupec}, D. and {H{\"u}tten}, M. and {Inada}, T. and {Iotov}, R. and {Ishio}, K. and {Iwamura}, Y. and {Jim{\'e}nez Mart{\'\i}nez}, I. and {Jormanainen}, J. and {Jouvin}, L. and {Kerszberg}, D. and {Kobayashi}, Y. and {Kubo}, H. and {Kushida}, J. and {Lamastra}, A. and {Lelas}, D. and {Leone}, F. and {Lindfors}, E. and {Linhoff}, L. and {Liodakis}, I. and {Lombardi}, S. and {Longo}, F. and {L{\'o}pez-Coto}, R. and {L{\'o}pez-Moya}, M. and {L{\'o}pez-Oramas}, A. and {Loporchio}, S. and {Lorini}, A. and {Machado de Oliveira Fraga}, B. and {Maggio}, C. and {Majumdar}, P. and {Makariev}, M. and {Mallamaci}, M. and {Maneva}, G. and {Manganaro}, M. and {Mannheim}, K. and {Mariotti}, M. and {Mart{\'\i}nez}, M. and {Mas Aguilar}, A. and {Mazin}, D. and {Menchiari}, S. and {Mender}, S. and {Mi{\'c}anovi{\'c}}, S. and {Miceli}, D. and {Miener}, T. and {Miranda}, J.~M. and {Mirzoyan}, R. and {Molina}, E. and {Mondal}, H.~A. and {Moralejo}, A. and {Morcuende}, D. and {Moreno}, V. and {Moretti}, E. and {Nakamori}, T. and {Nanci}, C. and {Nava}, L. and {Neustroev}, V. and {Nievas Rosillo}, M. and {Nigro}, C. and {Nilsson}, K. and {Nishijima}, K. and {Noda}, K. and {Nozaki}, S. and {Ohtani}, Y. and {Oka}, T. and {Otero-Santos}, J. and {Paiano}, S. and {Palatiello}, M. and {Paneque}, D. and {Paoletti}, R. and {Paredes}, J.~M. and {Pavleti{\'c}}, L. and {Pe{\~n}il}, P. and {Persic}, M. and {Pihet}, M. and {Prada Moroni}, P.~G. and {Prandini}, E. and {Priyadarshi}, C. and {Puljak}, I. and {Rhode}, W. and {Rib{\'o}}, M. and {Rico}, J. and {Righi}, C. and {Rugliancich}, A. and {Sahakyan}, N. and {Saito}, T. and {Sakurai}, S. and {Satalecka}, K. and {Saturni}, F.~G. and {Schleicher}, B. and {Schmidt}, K. and {Schmuckermaier}, F. and {Schubert}, J.~L. and {Schweizer}, T. and {Sitarek}, J. and {{\v{S}}nidari{\'c}}, I. and {Sobczynska}, D. and {Spolon}, A. and {Stamerra}, A. and {Stri{\v{s}}kovi{\'c}}, J. and {Strom}, D. and {Strzys}, M. and {Suda}, Y. and {Suri{\'c}}, T. and {Takahashi}, M. and {Takeishi}, R. and {Tavecchio}, F. and {Temnikov}, P. and {Terzi{\'c}}, T. and {Teshima}, M. and {Tosti}, L. and {Truzzi}, S. and {Tutone}, A. and {Ubach}, S. and {van Scherpenberg}, J. and {Vanzo}, G. and {Vazquez Acosta}, M. and {Ventura}, S. and {Verguilov}, V. and {Viale}, I. and {Vigorito}, C.~F. and {Vitale}, V.},
        title = "{A lower bound on intergalactic magnetic fields from time variability of 1ES 0229+200 from MAGIC and Fermi/LAT observations}",
      journal = {\aap},
         year = 2023,
        month = feb,
       volume = {670},
          eid = {A145},
        pages = {A145},
          doi = {10.1051/0004-6361/202244126},
archivePrefix = {arXiv},
       eprint = {2210.03321},
 primaryClass = {astro-ph.HE},
       adsurl = {https://ui.adsabs.harvard.edu/abs/2023A&A...670A.145A}
}

@ARTICLE{magic19_14C,
       author = {{MAGIC Collaboration} and {Acciari}, V.~A. and {Ansoldi}, S. and {Antonelli}, L.~A. and {Arbet Engels}, A. and {Baack}, D. and {Babi{\'c}}, A. and {Banerjee}, B. and {Barres de Almeida}, U. and {Barrio}, J.~A. and {Becerra Gonz{\'a}lez}, J. and {Bednarek}, W. and {Bellizzi}, L. and {Bernardini}, E. and {Berti}, A. and {Besenrieder}, J. and {Bhattacharyya}, W. and {Bigongiari}, C. and {Biland}, A. and {Blanch}, O. and {Bonnoli}, G. and {Bo{\v{s}}njak}, {\v{Z}}. and {Busetto}, G. and {Carosi}, A. and {Carosi}, R. and {Ceribella}, G. and {Chai}, Y. and {Chilingaryan}, A. and {Cikota}, S. and {Colak}, S.~M. and {Colin}, U. and {Colombo}, E. and {Contreras}, J.~L. and {Cortina}, J. and {Covino}, S. and {D'Amico}, G. and {D'Elia}, V. and {da Vela}, P. and {Dazzi}, F. and {de Angelis}, A. and {de Lotto}, B. and {Delfino}, M. and {Delgado}, J. and {Depaoli}, D. and {di Pierro}, F. and {di Venere}, L. and {Do Souto Espi{\~n}eira}, E. and {Dominis Prester}, D. and {Donini}, A. and {Dorner}, D. and {Doro}, M. and {Elsaesser}, D. and {Fallah Ramazani}, V. and {Fattorini}, A. and {Fern{\'a}ndez-Barral}, A. and {Ferrara}, G. and {Fidalgo}, D. and {Foffano}, L. and {Fonseca}, M.~V. and {Font}, L. and {Fruck}, C. and {Fukami}, S. and {Gallozzi}, S. and {Garc{\'\i}a L{\'o}pez}, R.~J. and {Garczarczyk}, M. and {Gasparyan}, S. and {Gaug}, M. and {Giglietto}, N. and {Giordano}, F. and {Godinovi{\'c}}, N. and {Green}, D. and {Guberman}, D. and {Hadasch}, D. and {Hahn}, A. and {Herrera}, J. and {Hoang}, J. and {Hrupec}, D. and {H{\"u}tten}, M. and {Inada}, T. and {Inoue}, S. and {Ishio}, K. and {Iwamura}, Y. and {Jouvin}, L. and {Kerszberg}, D. and {Kubo}, H. and {Kushida}, J. and {Lamastra}, A. and {Lelas}, D. and {Leone}, F. and {Lindfors}, E. and {Lombardi}, S. and {Longo}, F. and {L{\'o}pez}, M. and {L{\'o}pez-Coto}, R. and {L{\'o}pez-Oramas}, A. and {Loporchio}, S. and {Machado de Oliveira Fraga}, B. and {Maggio}, C. and {Majumdar}, P. and {Makariev}, M. and {Mallamaci}, M. and {Maneva}, G. and {Manganaro}, M. and {Mannheim}, K. and {Maraschi}, L. and {Mariotti}, M. and {Mart{\'\i}nez}, M. and {Masuda}, S. and {Mazin}, D. and {Mi{\'c}anovi{\'c}}, S. and {Miceli}, D. and {Minev}, M. and {Miranda}, J.~M. and {Mirzoyan}, R. and {Molina}, E. and {Moralejo}, A. and {Morcuende}, D. and {Moreno}, V. and {Moretti}, E. and {Munar-Adrover}, P. and {Neustroev}, V. and {Nigro}, C. and {Nilsson}, K. and {Ninci}, D. and {Nishijima}, K. and {Noda}, K. and {Nogu{\'e}s}, L. and {N{\"o}the}, M. and {Nozaki}, S. and {Paiano}, S. and {Palacio}, J. and {Palatiello}, M. and {Paneque}, D. and {Paoletti}, R. and {Paredes}, J.~M. and {Pe{\~n}il}, P. and {Peresano}, M. and {Persic}, M. and {Prada Moroni}, P.~G. and {Prandini}, E. and {Puljak}, I. and {Rhode}, W. and {Rib{\'o}}, M. and {Rico}, J. and {Righi}, C. and {Rugliancich}, A. and {Saha}, L. and {Sahakyan}, N. and {Saito}, T. and {Sakurai}, S. and {Satalecka}, K. and {Schmidt}, K. and {Schweizer}, T. and {Sitarek}, J. and {{\v{S}}nidari{\'c}}, I. and {Sobczynska}, D. and {Somero}, A. and {Stamerra}, A. and {Strom}, D. and {Strzys}, M. and {Suda}, Y. and {Suri{\'c}}, T. and {Takahashi}, M. and {Tavecchio}, F. and {Temnikov}, P. and {Terzi{\'c}}, T. and {Teshima}, M. and {Torres-Alb{\`a}}, N. and {Tosti}, L. and {Tsujimoto}, S. and {Vagelli}, V. and {van Scherpenberg}, J. and {Vanzo}, G. and {Vazquez Acosta}, M. and {Vigorito}, C.~F. and {Vitale}, V. and {Vovk}, I. and {Will}, M. and {Zari{\'c}}, D. and {Nava}, L.},
        title = "{Teraelectronvolt emission from the {\ensuremath{\gamma}}-ray burst GRB 190114C}",
      journal = {\nat},
         year = 2019,
        month = nov,
       volume = {575},
       number = {7783},
        pages = {455-458},
          doi = {10.1038/s41586-019-1750-x},
archivePrefix = {arXiv},
       eprint = {2006.07249},
 primaryClass = {astro-ph.HE},
       adsurl = {https://ui.adsabs.harvard.edu/abs/2019Natur.575..455M}
}

@ARTICLE{hess18,
       author = {{Abdalla}, H. and {Adam}, R. and {Aharonian}, F. and {Ait Benkhali}, F. and {Ang{\"u}ner}, E.~O. and {Arakawa}, M. and {Arcaro}, C. and {Armand}, C. and {Ashkar}, H. and {Backes}, M. and {Barbosa Martins}, V. and {Barnard}, M. and {Becherini}, Y. and {Berge}, D. and {Bernl{\"o}hr}, K. and {Bissaldi}, E. and {Blackwell}, R. and {B{\"o}ttcher}, M. and {Boisson}, C. and {Bolmont}, J. and {Bonnefoy}, S. and {Bregeon}, J. and {Breuhaus}, M. and {Brun}, F. and {Brun}, P. and {Bryan}, M. and {B{\"u}chele}, M. and {Bulik}, T. and {Bylund}, T. and {Capasso}, M. and {Caroff}, S. and {Carosi}, A. and {Casanova}, S. and {Cerruti}, M. and {Chand}, T. and {Chandra}, S. and {Chen}, A. and {Colafrancesco}, S. and {Cury{\l}o}, M. and {Davids}, I.~D. and {Deil}, C. and {Devin}, J. and {deWilt}, P. and {Dirson}, L. and {Djannati-Ata{\"\i}}, A. and {Dmytriiev}, A. and {Donath}, A. and {Doroshenko}, V. and {Dyks}, J. and {Egberts}, K. and {Emery}, G. and {Ernenwein}, J. -P. and {Eschbach}, S. and {Feijen}, K. and {Fegan}, S. and {Fiasson}, A. and {Fontaine}, G. and {Funk}, S. and {F{\"u}{\ss}ling}, M. and {Gabici}, S. and {Gallant}, Y.~A. and {Gat{\'e}}, F. and {Giavitto}, G. and {Giunti}, L. and {Glawion}, D. and {Glicenstein}, J.~F. and {Gottschall}, D. and {Grondin}, M. -H. and {Hahn}, J. and {Haupt}, M. and {Heinzelmann}, G. and {Henri}, G. and {Hermann}, G. and {Hinton}, J.~A. and {Hofmann}, W. and {Hoischen}, C. and {Holch}, T.~L. and {Holler}, M. and {Horns}, D. and {Huber}, D. and {Iwasaki}, H. and {Jamrozy}, M. and {Jankowsky}, D. and {Jankowsky}, F. and {Jardin-Blicq}, A. and {Jung-Richardt}, I. and {Kastendieck}, M.~A. and {Katarzy{\'n}ski}, K. and {Katsuragawa}, M. and {Katz}, U. and {Khangulyan}, D. and {Kh{\'e}lifi}, B. and {King}, J. and {Klepser}, S. and {Klu{\'z}niak}, W. and {Komin}, Nu. and {Kosack}, K. and {Kostunin}, D. and {Kreter}, M. and {Lamanna}, G. and {Lemi{\`e}re}, A. and {Lemoine-Goumard}, M. and {Lenain}, J. -P. and {Leser}, E. and {Levy}, C. and {Lohse}, T. and {Lypova}, I. and {Mackey}, J. and {Majumdar}, J. and {Malyshev}, D. and {Marandon}, V. and {Marcowith}, A. and {Mares}, A. and {Mariaud}, C. and {Mart{\'\i}-Devesa}, G. and {Marx}, R. and {Maurin}, G. and {Meintjes}, P.~J. and {Mitchell}, A.~M.~W. and {Moderski}, R. and {Mohamed}, M. and {Mohrmann}, L. and {Moore}, C. and {Moulin}, E. and {Muller}, J. and {Murach}, T. and {Nakashima}, S. and {de Naurois}, M. and {Ndiyavala}, H. and {Niederwanger}, F. and {Niemiec}, J. and {Oakes}, L. and {O'Brien}, P. and {Odaka}, H. and {Ohm}, S. and {de Ona Wilhelmi}, E. and {Ostrowski}, M. and {Oya}, I. and {Panter}, M. and {Parsons}, R.~D. and {Perennes}, C. and {Petrucci}, P. -O. and {Peyaud}, B. and {Piel}, Q. and {Pita}, S. and {Poireau}, V. and {Priyana Noel}, A. and {Prokhorov}, D.~A. and {Prokoph}, H. and {P{\"u}hlhofer}, G. and {Punch}, M. and {Quirrenbach}, A. and {Raab}, S. and {Rauth}, R. and {Reimer}, A. and {Reimer}, O. and {Remy}, Q. and {Renaud}, M. and {Rieger}, F. and {Rinchiuso}, L. and {Romoli}, C. and {Rowell}, G. and {Rudak}, B. and {Ruiz-Velasco}, E. and {Sahakian}, V. and {Sailer}, S. and {Saito}, S. and {Sanchez}, D.~A. and {Santangelo}, A. and {Sasaki}, M. and {Schlickeiser}, R. and {Sch{\"u}ssler}, F. and {Schulz}, A. and {Schutte}, H.~M. and {Schwanke}, U. and {Schwemmer}, S. and {Seglar-Arroyo}, M. and {Senniappan}, M. and {Seyffert}, A.~S. and {Shafi}, N. and {Shiningayamwe}, K. and {Simoni}, R. and {Sinha}, A. and {Sol}, H. and {Specovius}, A. and {Spir-Jacob}, M. and {Stawarz}, {\L}. and {Steenkamp}, R. and {Stegmann}, C. and {Steppa}, C. and {Takahashi}, T. and {Tavernier}, T. and {Taylor}, A.~M. and {Terrier}, R. and {Tiziani}, D. and {Tluczykont}, M. and {Trichard}, C. and {Tsirou}, M. and {Tsuji}, N. and {Tuffs}, R. and {Uchiyama}, Y. and {van der Walt}, D.~J. and {van Eldik}, C. and {van Rensburg}, C. and {van Soelen}, B. and {Vasileiadis}, G. and {Veh}, J. and {Venter}, C. and {Vincent}, P. and {Vink}, J. and {V{\"o}lk}, H.~J. and {Vuillaume}, T. and {Wadiasingh}, Z. and {Wagner}, S.~J. and {White}, R. and {Wierzcholska}, A. and {Yang}, R. and {Yoneda}, H. and {Zacharias}, M. and {Zanin}, R. and {Zdziarski}, A.~A. and {Zech}, A. and {Ziegler}, A. and {Zorn}, J. and {{\.Z}ywucka}, N. and {de Palma}, F. and {Axelsson}, M. and {Roberts}, O.~J.},
        title = "{A very-high-energy component deep in the {\ensuremath{\gamma}}-ray burst afterglow}",
      journal = {\nat},
         year = 2019,
        month = nov,
       volume = {575},
       number = {7783},
        pages = {464-467},
          doi = {10.1038/s41586-019-1743-9},
archivePrefix = {arXiv},
       eprint = {1911.08961},
 primaryClass = {astro-ph.HE},
       adsurl = {https://ui.adsabs.harvard.edu/abs/2019Natur.575..464A}
}

@ARTICLE{Plaga95,
       author = {{Plaga}, R.},
        title = "{Detecting intergalactic magnetic fields using time delays in pulses of {\ensuremath{\gamma}}-rays}",
      journal = {Nature},
         year = 1995,
        month = mar,
       volume = {374},
       number = {6521},
        pages = {430-432},
          doi = {10.1038/374430a0},
       adsurl = {https://ui.adsabs.harvard.edu/abs/1995Natur.374..430P}
}

@ARTICLE{Razzaque04,
       author = {{Razzaque}, Soebur and {M{\'e}sz{\'a}ros}, Peter and {Zhang}, Bing},
        title = "{GeV and Higher Energy Photon Interactions in Gamma-Ray Burst Fireballs and Surroundings}",
      journal = {\apj},
         year = 2004,
        month = oct,
       volume = {613},
       number = {2},
        pages = {1072-1078},
          doi = {10.1086/423166},
archivePrefix = {arXiv},
       eprint = {astro-ph/0404076},
 primaryClass = {astro-ph},
       adsurl = {https://ui.adsabs.harvard.edu/abs/2004ApJ...613.1072R}
}

@ARTICLE{Takahashi11,
       author = {{Takahashi}, Keitaro and {Inoue}, Susumu and {Ichiki}, Kiyotomo and {Nakamura}, Takashi},
        title = "{Probing early cosmic magnetic fields through pair echoes from high-redshift GRBs}",
      journal = {Mon. Not. R. Astron. Soc.},
         year = 2011,
        month = feb,
       volume = {410},
       number = {4},
        pages = {2741-2748},
          doi = {10.1111/j.1365-2966.2010.17639.x},
archivePrefix = {arXiv},
       eprint = {1007.5363},
 primaryClass = {astro-ph.HE},
       adsurl = {https://ui.adsabs.harvard.edu/abs/2011MNRAS.410.2741T}
}

@ARTICLE{Murase08,
       author = {{Murase}, Kohta and {Takahashi}, Keitaro and {Inoue}, Susumu and {Ichiki}, Kiyomoto and {Nagataki}, Shigehiro},
        title = "{Probing Intergalactic Magnetic Fields in the GLAST Era through Pair Echo Emission from TeV Blazars}",
      journal = {\apjl},
         year = 2008,
        month = oct,
       volume = {686},
       number = {2},
        pages = {L67},
          doi = {10.1086/592997},
archivePrefix = {arXiv},
       eprint = {0806.2829},
 primaryClass = {astro-ph},
       adsurl = {https://ui.adsabs.harvard.edu/abs/2008ApJ...686L..67M}
}

@ARTICLE{Murase09,
       author = {{Murase}, Kohta and {Zhang}, Bing and {Takahashi}, Keitaro and {Nagataki}, Shigehiro},
        title = "{Possible effects of pair echoes on gamma-ray burst afterglow emission}",
      journal = {\mnras},
         year = 2009,
        month = jul,
       volume = {396},
       number = {4},
        pages = {1825-1832},
          doi = {10.1111/j.1365-2966.2009.14704.x},
archivePrefix = {arXiv},
       eprint = {0812.0124},
 primaryClass = {astro-ph},
       adsurl = {https://ui.adsabs.harvard.edu/abs/2009MNRAS.396.1825M}
}

@ARTICLE{wang20,
       author = {{Wang}, Ze-Rui and {Xi}, Shao-Qiang and {Liu}, Ruo-Yu and {Xue}, Rui and {Wang}, Xiang-Yu},
        title = "{Constraints on the intergalactic magnetic field from {\ensuremath{\gamma}} -ray observations of GRB 190114C}",
      journal = {\prd},
         year = 2020,
        month = apr,
       volume = {101},
       number = {8},
          eid = {083004},
        pages = {083004},
          doi = {10.1103/PhysRevD.101.083004},
archivePrefix = {arXiv},
       eprint = {2001.01186},
 primaryClass = {astro-ph.HE},
       adsurl = {https://ui.adsabs.harvard.edu/abs/2020PhRvD.101h3004W}
}

@ARTICLE{dzhatdoev20,
       author = {{Dzhatdoev}, T.~A. and {Podlesnyi}, E.~I. and {Vaiman}, I.~A.},
        title = "{Can we constrain the extragalactic magnetic field from very high energy observations of GRB 190114C?}",
      journal = {\prd},
         year = 2020,
        month = dec,
       volume = {102},
       number = {12},
          eid = {123017},
        pages = {123017},
          doi = {10.1103/PhysRevD.102.123017},
archivePrefix = {arXiv},
       eprint = {2002.06918},
 primaryClass = {astro-ph.HE},
       adsurl = {https://ui.adsabs.harvard.edu/abs/2020PhRvD.102l3017D}
}

@ARTICLE{Vovk23,
       author = {{Vovk}, Ievgen},
        title = "{Search of the pair echo signatures in the high-energy light curve of GRB190114C}",
      journal = {\prd},
         year = 2023,
        month = feb,
       volume = {107},
       number = {4},
          eid = {043020},
        pages = {043020},
          doi = {10.1103/PhysRevD.107.043020},
archivePrefix = {arXiv},
       eprint = {2301.08432},
 primaryClass = {astro-ph.HE},
       adsurl = {https://ui.adsabs.harvard.edu/abs/2023PhRvD.107d3020V}
}

@ARTICLE{lhaaso23,
       author = {{LHAASO Collaboration} and {Cao}, Z. and {Aharonian}, F. and {An}, Q. and {Axikegu}, A. and {Bai}, L.~X. and {Bai}, Y.~X. and {Bao}, Y.~W. and {Bastieri}, D. and {Bi}, X.~J. and {Bi}, Y.~J. and {Cai}, J.~T. and {Cao}, Q. and {Cao}, W.~Y. and {Cao}, Z. and {Chang}, J. and {Chang}, J.~F. and {Chen}, E.~S. and {Chen}, L. and {Chen}, L. and {Chen}, L. and {Chen}, M.~J. and {Chen}, M.~L. and {Chen}, Q.~H. and {Chen}, S.~H. and {Chen}, S.~Z. and {Chen}, T.~L. and {Chen}, Y. and {Cheng}, H.~L. and {Cheng}, N. and {Cheng}, Y.~D. and {Cui}, S.~W. and {Cui}, X.~H. and {Cui}, Y.~D. and {Dai}, B.~Z. and {Dai}, H.~L. and {Danzengluobu}, D. and {Della Volpe}, D. and {Dong}, X.~Q. and {Duan}, K.~K. and {Fan}, J.~H. and {Fan}, Y.~Z. and {Fang}, J. and {Fang}, K. and {Feng}, C.~F. and {Feng}, L. and {Feng}, S.~H. and {Feng}, X.~T. and {Feng}, Y.~L. and {Gao}, B. and {Gao}, C.~D. and {Gao}, L.~Q. and {Gao}, Q. and {Gao}, W. and {Gao}, W.~K. and {Ge}, M.~M. and {Geng}, L.~S. and {Gong}, G.~H. and {Gou}, Q.~B. and {Gu}, M.~H. and {Guo}, F.~L. and {Guo}, X.~L. and {Guo}, Y.~Q. and {Guo}, Y.~Y. and {Han}, Y.~A. and {He}, H.~H. and {He}, H.~N. and {He}, J.~Y. and {He}, X.~B. and {He}, Y. and {Heller}, M. and {Hor}, Y.~K. and {Hou}, B.~W. and {Hou}, C. and {Hou}, X. and {Hu}, H.~B. and {Hu}, Q. and {Hu}, S.~C. and {Huang}, D.~H. and {Huang}, T.~Q. and {Huang}, W.~J. and {Huang}, X.~T. and {Huang}, Z.~C. and {Ji}, X.~L. and {Jia}, H.~Y. and {Jia}, K. and {Jiang}, K. and {Jiang}, X.~W. and {Jiang}, Z.~J. and {Jin}, M. and {Kang}, M.~M. and {Ke}, T. and {Kuleshov}, D. and {Kurinov}, K. and {Li}, B.~B. and {Li}, C. and {Li}, C. and {Li}, D. and {Li}, F. and {Li}, H.~B. and {Li}, H.~C. and {Li}, H.~Y. and {Li}, J. and {Li}, J. and {Li}, J. and {Li}, K. and {Li}, W.~L. and {Li}, W.~L. and {Li}, X.~R. and {Li}, X. and {Li}, Y.~Z. and {Li}, Z. and {Li}, Z. and {Liang}, E.~W. and {Liang}, Y.~F. and {Lin}, S.~J. and {Liu}, B. and {Liu}, C. and {Liu}, D. and {Liu}, H. and {Liu}, H.~D. and {Liu}, J. and {Liu}, J.~L. and {Liu}, J.~L. and {Liu}, J.~S. and {Liu}, J.~Y. and {Liu}, M.~Y. and {Liu}, R.~Y. and {Liu}, S.~M. and {Liu}, W. and {Liu}, Y. and {Liu}, Y.~N. and {Long}, W.~J. and {Lu}, R. and {Luo}, Q. and {Lv}, H.~K. and {Ma}, B.~Q. and {Ma}, L.~L. and {Ma}, X.~H. and {Mao}, J.~R. and {Min}, Z. and {Mitthumsiri}, W. and {Nan}, Y.~C. and {Ou}, Z.~W. and {Pang}, B.~Y. and {Pattarakijwanich}, P. and {Pei}, Z.~Y. and {Qi}, M.~Y. and {Qi}, Y.~Q. and {Qiao}, B.~Q. and {Qin}, J.~J. and {Ruffolo}, D. and {Saiz}, A. and {Shao}, C.~Y. and {Shao}, L. and {Shchegolev}, O. and {Sheng}, X.~D. and {Song}, H.~C. and {Stenkin}, Y.~V. and {Stepanov}, V. and {Su}, Y. and {Sun}, Q.~N. and {Sun}, X.~N. and {Sun}, Z.~B. and {Tam}, P.~H.~T. and {Tang}, Z.~B. and {Tian}, W.~W. and {Wang}, C. and {Wang}, C.~B. and {Wang}, G.~W. and {Wang}, H.~G. and {Wang}, H.~H. and {Wang}, J.~C. and {Wang}, J.~S. and {Wang}, K. and {Wang}, L.~P. and {Wang}, L.~Y. and {Wang}, P.~H. and {Wang}, R. and {Wang}, W. and {Wang}, X.~G. and {Wang}, Y.~D. and {Wang}, Y.~J. and {Wang}, Z.~H. and {Wang}, Z.~X. and {Wang}, Z. and {Wei}, D.~M. and {Wei}, J.~J. and {Wei}, Y.~J. and {Wen}, T. and {Wu}, C.~Y. and {Wu}, H.~R. and {Wu}, S. and {Wu}, X.~F. and {Wu}, Y.~S. and {Xi}, S.~Q. and {Xia}, J. and {Xia}, J.~J. and {Xiang}, G.~M. and {Xiao}, D.~X. and {Xiao}, G. and {Xin}, G.~G. and {Xin}, Y.~L. and {Xing}, Y. and {Xiong}, Z. and {Xu}, D.~L. and {Xu}, R.~F. and {Xu}, R.~X. and {Xue}, L. and {Yan}, D.~H. and {Yan}, J.~Z. and {Yan}, T. and {Yang}, C.~W. and {Yang}, F. and {Yang}, F.~F. and {Yang}, H.~W. and {Yang}, J.~Y. and {Yang}, L.~L. and {Yang}, M.~J. and {Yang}, R.~Z. and {Yang}, S.~B. and {Yao}, Y.~H. and {Ye}, Y.~M. and {Yin}, L.~Q. and {Yin}, N. and {You}, X.~H. and {You}, Z.~Y. and {Yu}, Y.~H. and {Yuan}, Q. and {Yue}, H. and {Zeng}, H.~D. and {Zeng}, T.~X. and {Zeng}, W. and {Zeng}, Z.~K. and {Zhang}, B. and {Zhang}, B.~B. and {Zhang}, F. and {Zhang}, H.~M. and {Zhang}, H.~Y. and {Zhang}, J.~L. and {Zhang}, L.~X. and {Zhang}, L. and {Zhang}, P.~F. and {Zhang}, P.~P. and {Zhang}, R. and {Zhang}, S.~B. and {Zhang}, S.~R. and {Zhang}, S.~S. and {Zhang}, X. and {Zhang}, X.~P. and {Zhang}, Y.~F. and {Zhang}, Y. and {Zhang}, Y. and {Zhao}, B. and {Zhao}, J. and {Zhao}, L. and {Zhao}, L.~Z. and {Zhao}, S.~P. and {Zheng}, F. and {Zhou}, B. and {Zhou}, H. and {Zhou}, J.~N. and {Zhou}, P. and {Zhou}, R. and {Zhou}, X.~X. and {Zhu}, C.~G. and {Zhu}, F.~R. and {Zhu}, H. and {Zhu}, K.~J. and {Zuo}, X.},
        title = "{A tera-electron volt afterglow from a narrow jet in an extremely bright gamma-ray burst.}",
      journal = {Science},
         year = 2023,
        month = jun,
       volume = {380},
       number = {6652},
        pages = {1390-1396},
          doi = {10.1126/science.adg9328},
archivePrefix = {arXiv},
       eprint = {2306.06372},
 primaryClass = {astro-ph.HE},
       adsurl = {https://ui.adsabs.harvard.edu/abs/2023Sci...380.1390L}
}

@ARTICLE{Huang23,
       author = {{Huang}, Yi-Yun and {Dai}, Cui-Yuan and {Zhang}, Hai-Ming and {Liu}, Ruo-Yu and {Wang}, Xiang-Yu},
        title = "{Constraints on the Intergalactic Magnetic Field Strength from {\ensuremath{\gamma}}-Ray Observations of GRB 221009A}",
      journal = {\apjl},
         year = 2023,
        month = sep,
       volume = {955},
       number = {1},
          eid = {L10},
        pages = {L10},
          doi = {10.3847/2041-8213/acf66a},
archivePrefix = {arXiv},
       eprint = {2306.05970},
 primaryClass = {astro-ph.HE},
       adsurl = {https://ui.adsabs.harvard.edu/abs/2023ApJ...955L..10H}
}

@ARTICLE{Dzhatdoev24,
       author = {{Dzhatdoev}, Timur A. and {Podlesnyi}, Egor I. and {Rubtsov}, Grigory I.},
        title = "{First constraints on the strength of the extragalactic magnetic field from {\ensuremath{\gamma}}-ray observations of GRB 221009A}",
      journal = {\mnras},
         year = 2024,
        month = jan,
       volume = {527},
       number = {1},
        pages = {L95-L102},
          doi = {10.1093/mnrasl/slad142},
archivePrefix = {arXiv},
       eprint = {2306.05347},
 primaryClass = {astro-ph.HE},
       adsurl = {https://ui.adsabs.harvard.edu/abs/2024MNRAS.527L..95D}
}

@ARTICLE{Vovk23b,
       author = {{Vovk}, Ie. and {Korochkin}, A. and {Neronov}, A. and {Semikoz}, D.},
        title = "{Constraint on intergalactic magnetic field from Fermi/LAT observations of the ``pair echo'' of GRB 221009A}",
      journal = {arXiv e-prints},
         year = 2023,
        month = jun,
          eid = {arXiv:2306.07672},
        pages = {arXiv:2306.07672},
          doi = {10.48550/arXiv.2306.07672},
archivePrefix = {arXiv},
       eprint = {2306.07672},
 primaryClass = {astro-ph.HE},
       adsurl = {https://ui.adsabs.harvard.edu/abs/2023arXiv230607672V}
}

@ARTICLE{crpropa3,
       author = {{Alves Batista}, Rafael and {Becker Tjus}, Julia and {D{\"o}rner}, Julien and {Dundovic}, Andrej and {Eichmann}, Bj{\"o}rn and {Frie}, Antonius and {Heiter}, Christopher and {Hoerbe}, Mario R. and {Kampert}, Karl-Heinz and {Merten}, Lukas and {M{\"u}ller}, Gero and {Reichherzer}, Patrick and {Saveliev}, Andrey and {Schlegel}, Leander and {Sigl}, G{\"u}nter and {van Vliet}, Arjen and {Winchen}, Tobias},
        title = "{CRPropa 3.2 - an advanced framework for high-energy particle propagation in extragalactic and galactic spaces}",
      journal = {\jcap},
         year = 2022,
        month = sep,
       volume = {2022},
       number = {9},
          eid = {035},
        pages = {035},
          doi = {10.1088/1475-7516/2022/09/035},
archivePrefix = {arXiv},
       eprint = {2208.00107},
 primaryClass = {astro-ph.HE},
       adsurl = {https://ui.adsabs.harvard.edu/abs/2022JCAP...09..035A}
}

@ARTICLE{Veres22,
       author = {{Veres}, P. and {Burns}, E. and {Bissaldi}, E. and {Lesage}, S. and {Roberts}, O. and {Fermi GBM Team}},
        title = "{GRB 221009A: Fermi GBM detection of an extraordinarily bright GRB}",
      journal = {GRB Coordinates Network},
         year = 2022,
        month = oct,
       volume = {32636},
        pages = {1},
       adsurl = {https://ui.adsabs.harvard.edu/abs/2022GCN.32636....1V}
}

@ARTICLE{Postigo22,
       author = {{de Ugarte Postigo}, A. and {Izzo}, L. and {Pugliese}, G. and {Xu}, D. and {Schneider}, B. and {Fynbo}, J.~P.~U. and {Tanvir}, N.~R. and {Malesani}, D.~B. and {Saccardi}, A. and {Kann}, D.~A. and {Wiersema}, K. and {Gompertz}, B.~P. and {Thoene}, C.~C. and {Levan}, A.~J. and {Stargate Collaboration}},
        title = "{GRB 221009A: Redshift from X-shooter/VLT}",
      journal = {GRB Coordinates Network},
         year = 2022,
        month = oct,
       volume = {32648},
        pages = {1},
       adsurl = {https://ui.adsabs.harvard.edu/abs/2022GCN.32648....1D}
}

@ARTICLE{Bissaldi22,
       author = {{Bissaldi}, E. and {Omodei}, N. and {Kerr}, M. and {Fermi-LAT Team}},
        title = "{GRB 221009A or Swift J1913.1+1946: Fermi-LAT detection}",
      journal = {GRB Coordinates Network},
         year = 2022,
        month = oct,
       volume = {32637},
        pages = {1},
       adsurl = {https://ui.adsabs.harvard.edu/abs/2022GCN.32637....1B}
}

@ARTICLE{Dichiara22,
       author = {{Dichiara}, S. and {Gropp}, J.~D. and {Kennea}, J.~A. and {Kuin}, N.~P.~M. and {Lien}, A.~Y. and {Marshall}, F.~E. and {Tohuvavohu}, A. and {Williams}, M.~A. and {Neil Gehrels Swift Observatory Team}},
        title = "{Swift J1913.1+1946 a new bright hard X-ray and optical transient}",
      journal = {GRB Coordinates Network},
         year = 2022,
        month = oct,
       volume = {32632},
        pages = {1},
       adsurl = {https://ui.adsabs.harvard.edu/abs/2022GCN.32632....1D}
}

@ARTICLE{Xin22,
       author = {{Ma}, Xin-Hua and {Bi}, Yu-Jiang and {Cao}, Zhen and {Chen}, Ming-Jun and {Chen}, Song-Zhan and {Cheng}, Yao-Dong and {Gong}, Guang-Hua and {Gu}, Min-Hao and {He}, Hui-Hai and {Hou}, Chao and {Huang}, Wen-Hao and {Huang}, Xing-Tao and {Liu}, Cheng and {Shchegolev}, Oleg and {Sheng}, Xiang-Dong and {Stenkin}, Yuri and {Wu}, Chao-Yong and {Wu}, Han-Rong and {Wu}, Sha and {Xiao}, Gang and {Yao}, Zhi-Guo and {Zhang}, Shou-Shan and {Zhang}, Yi and {Zuo}, Xiong},
        title = "{Chapter 1 LHAASO Instruments and Detector technology}",
      journal = {Chinese Physics C},
         year = 2022,
        month = mar,
       volume = {46},
       number = {3},
          eid = {030001},
        pages = {030001},
          doi = {10.1088/1674-1137/ac3fa6},
       adsurl = {https://ui.adsabs.harvard.edu/abs/2022ChPhC..46c0001M}
}

@ARTICLE{Cao19,
       author = {{Cao}, Zhen and {Chen}, Ming-Jun and {Chen}, Song-Zhan Hu, Hong-Bo and {Liu}, Cheng and {Liu}, Ye and {Ma}, Ling-Ling and {Ma}, Xin-Hua and {Sheng}, Xiang-Dong and {Wu}, Han-Rong and {Xiao}, Gang and {Yao}, Zhi-Guo and {Yin}, Li-Qiao and {Zha}, Min and {Zhang}, Shou-Shan and {LHAASO Collaboration}},
        title = "{Introduction to Large High Altitude Air Shower Observatory (LHAASO)}",
      journal = {Chinese Astronomy and Astrophysics},
         year = 2019,
        month = oct,
       volume = {43},
       number = {4},
        pages = {457-478},
          doi = {10.1016/j.chinastron.2019.11.001},
       adsurl = {https://ui.adsabs.harvard.edu/abs/2019ChA&A..43..457C}
}

@ARTICLE{franceschini08,
       author = {{Franceschini}, A. and {Rodighiero}, G. and {Vaccari}, M.},
        title = "{Extragalactic optical-infrared background radiation, its time evolution and the cosmic photon-photon opacity}",
      journal = {\aap},
         year = 2008,
        month = sep,
       volume = {487},
       number = {3},
        pages = {837-852},
          doi = {10.1051/0004-6361:200809691},
archivePrefix = {arXiv},
       eprint = {0805.1841},
 primaryClass = {astro-ph},
       adsurl = {https://ui.adsabs.harvard.edu/abs/2008A&A...487..837F}
}

@ARTICLE{Dominguez11,
       author = {{Dom{\'\i}nguez}, A. and {Primack}, J.~R. and {Rosario}, D.~J. and {Prada}, F. and {Gilmore}, R.~C. and {Faber}, S.~M. and {Koo}, D.~C. and {Somerville}, R.~S. and {P{\'e}rez-Torres}, M.~A. and {P{\'e}rez-Gonz{\'a}lez}, P. and {Huang}, J. -S. and {Davis}, M. and {Guhathakurta}, P. and {Barmby}, P. and {Conselice}, C.~J. and {Lozano}, M. and {Newman}, J.~A. and {Cooper}, M.~C.},
        title = "{Extragalactic background light inferred from AEGIS galaxy-SED-type fractions}",
      journal = {\mnras},
         year = 2011,
        month = feb,
       volume = {410},
       number = {4},
        pages = {2556-2578},
          doi = {10.1111/j.1365-2966.2010.17631.x},
archivePrefix = {arXiv},
       eprint = {1007.1459},
 primaryClass = {astro-ph.CO},
       adsurl = {https://ui.adsabs.harvard.edu/abs/2011MNRAS.410.2556D}
}

@ARTICLE{Zhou24,
       author = {{Zhou}, Lin and {Zou}, Yuan-Chuan},
        title = "{Determining the viewing angle from TeV light curve of GRB 221009A}",
      journal = {\mnras},
         year = 2024,
        month = aug,
       volume = {532},
       number = {2},
        pages = {2189-2195},
          doi = {10.1093/mnras/stae1644},
archivePrefix = {arXiv},
       eprint = {2407.03743},
 primaryClass = {astro-ph.HE},
       adsurl = {https://ui.adsabs.harvard.edu/abs/2024MNRAS.532.2189Z}
}

%\newpage
\clearpage

\appendix

\section{CRPropa Simulations Details} \label{app:CRPROPA}

\begin{figure*}[ht]
    \centering
    \includegraphics[width=0.49\linewidth]{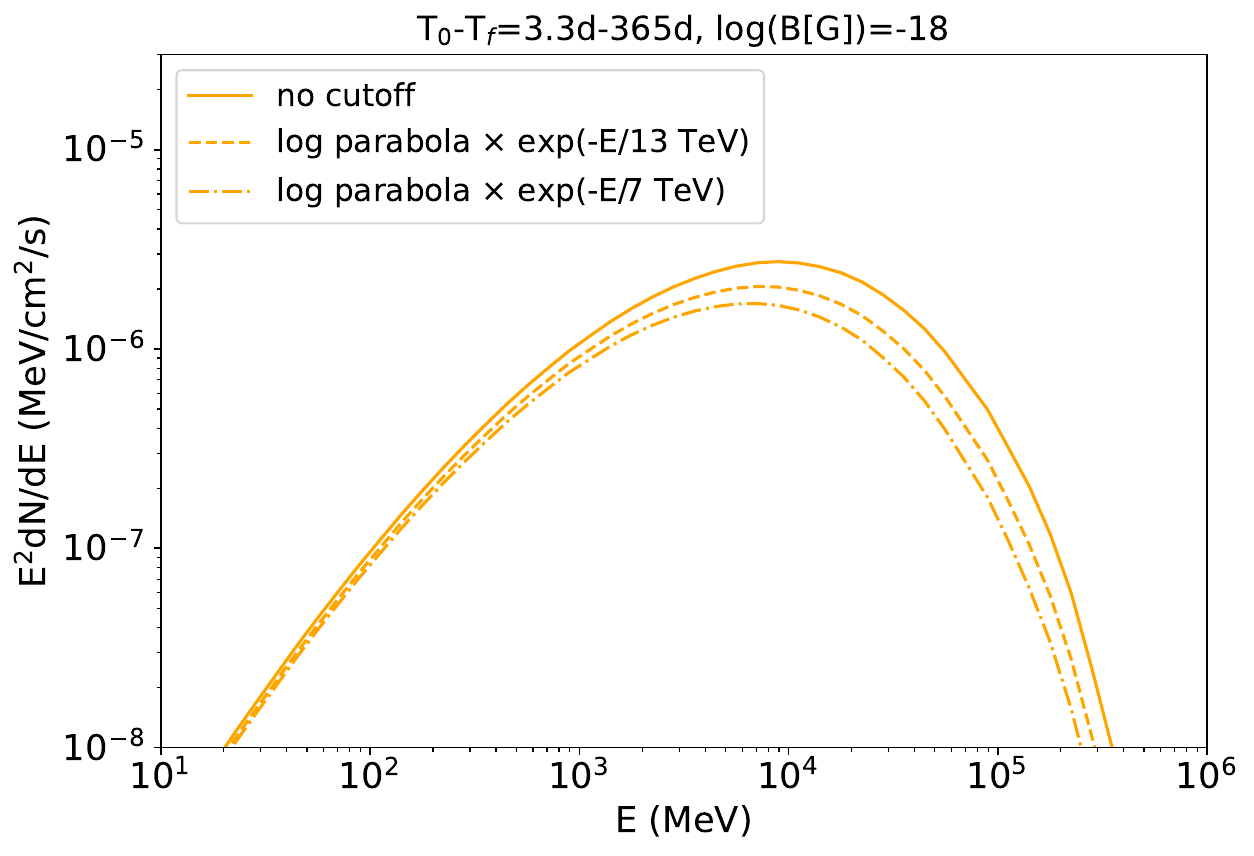}
    \includegraphics[width=0.49\linewidth]{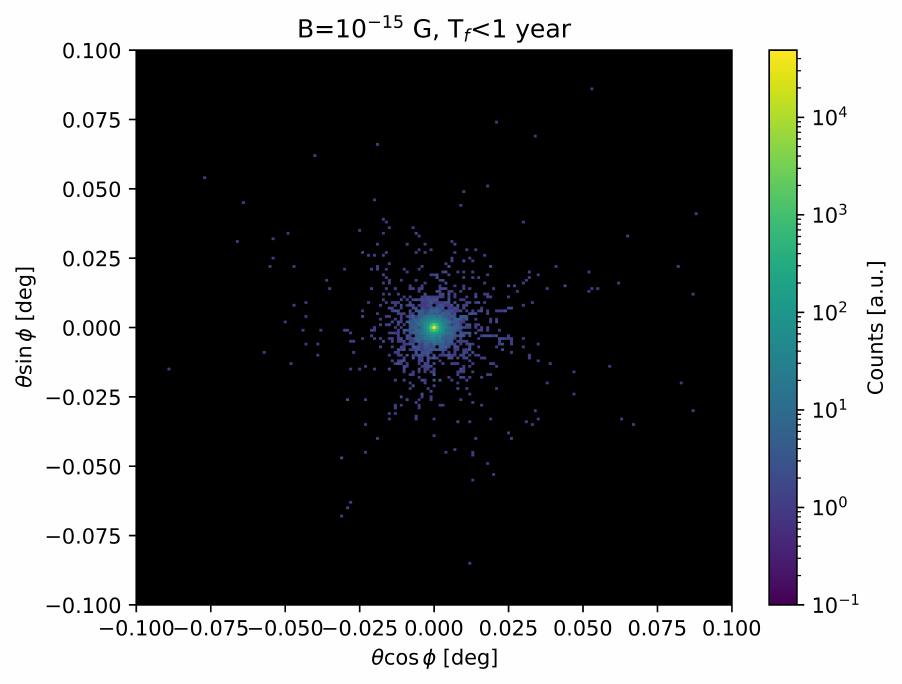}
    \caption{\label{fig:cascade_and_map} Left: Pair-echo SEDs comparison for $\log_{10}(B/\mathrm{G})=-18$ and the three different VHE intrinsic spectra used in the simulations. Right: 2D pair-echo distribution for $\log_{10}(B/\mathrm{G})=-15$ (the strongest magnetic field strength tested) and $E>10$~MeV. As a reference, for average reconstruction quality events the smallest LAT PSF is above $10$~GeV, with a $68\%$ containment within 0.1$^\circ$ \cite{4FGL}.}
\end{figure*}

\begin{figure*}[ht]
    \centering
    \includegraphics[width=0.49\linewidth]{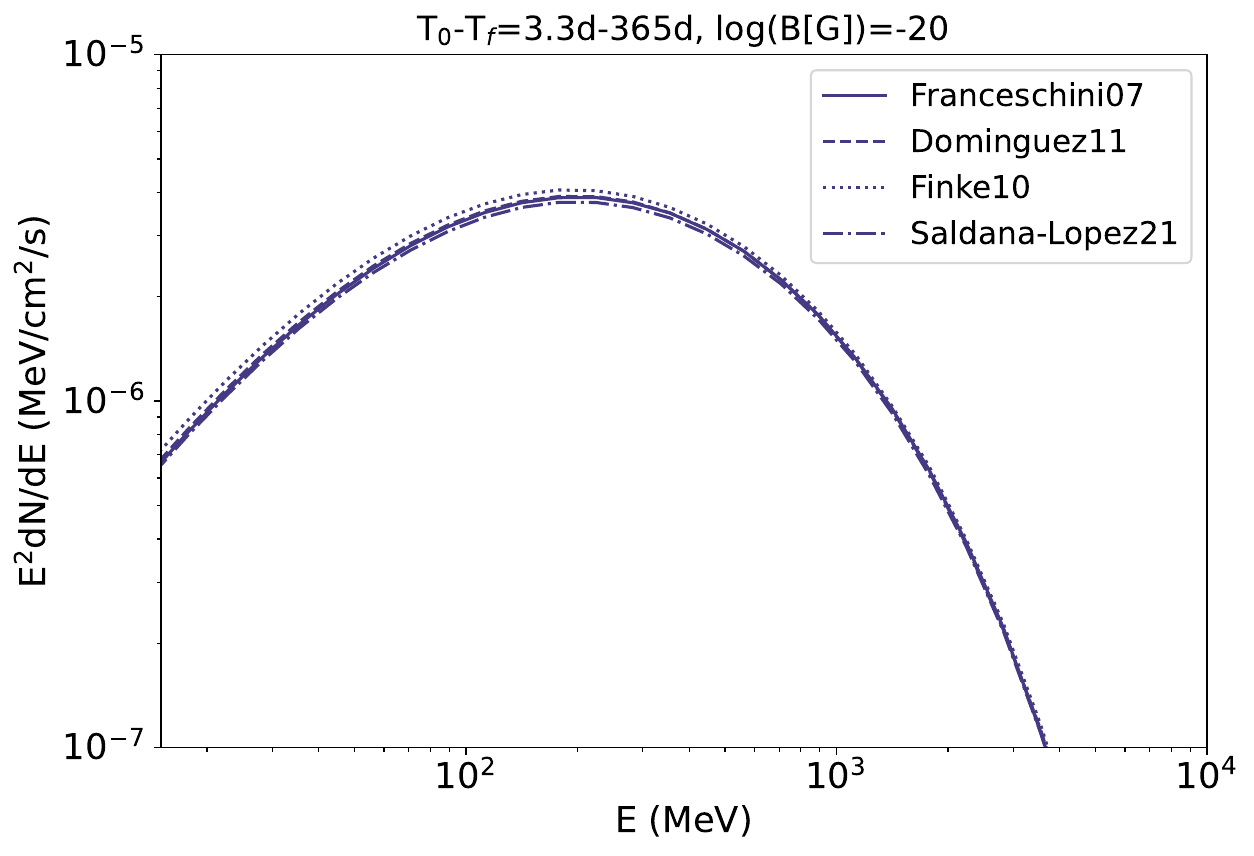}
    \caption{\label{fig:ebl_test} Comparison of pair-echo SEDs for $\log_{10}(B/\mathrm{G})=-20$ and a log-parabola VHE intrinsic spectrum, using different EBL models (see text for details).}
\end{figure*}

As explained in the main body of the paper, in the simulation setup, the source is positioned at the center of a sphere with a radius equal to the comoving source distance. Very-high-energy (VHE) photons are injected within a cone, and all relevant energy-loss mechanisms are accounted for, including pair production and inverse Compton (IC) scattering on the cosmic microwave background (CMB). The extragalactic background light (EBL) model from \cite{Franceschini08} is used in the simulations. 
A pair-echo photon with energy larger than 10 MeV that hits the sphere is considered as detected.

We assume an IGMF with a turbulent zero-mean Gaussian random field and a Kolmogorov spectrum.
It is defined in Fourier space,
transformed into real space, and then projected onto a uniformly
spaced cubic grid with $N = 100^3$ cells of 50\,Mpc size. The minimum and maximum scale lengths in which the magnetic field
are defined are 1\,Mpc and 25\,Mpc, respectively, resulting in a correlation length of $\ell_B \sim5\,$Mpc.
The cubic cell is then periodically repeated to cover the distance between the source and Earth.
We conduct the simulations for different values of the root mean square of the IGMF field strength between $10^{-20}\,$G and $10^{-15}\,$G.
For each tested field and VHE spectrum with different cutoff energies, we inject $2\times10^5$ photons. 
For the two strongest IGMF strengths ($\log_{10}(B/ \mathrm{G})=-15.5$ and $-15$), we inject $4\times10^5$ and $8\times10^5$ photons, respectively, due to the strong time dilution of the pair-echo signal.
All particles are traced with a step size of at least 10$^{-5}$ pc which enables us to reproduce time delays with an accuracy better than 20 minutes.

To evaluate the pair-echo SED within a certain time interval $\Delta T$, we consider an observer perfectly aligned with the jet axis\footnote{The assumption was justified by existing estimates, see Table 3 in \citep{Zhou24}.} and select all cascade photons arriving with a time delay $< \Delta T$~\cite{DaVela23}. In the left panel of Fig.~\ref{fig:cascade_and_map}, we show an example of the SED derived in the time interval 3.3-365\,days for the three different VHE intrinsic spectra tested; a log parabola and a log parabola multiplied with two different exponential cutoff energies. 

In principle we also need to consider the spatial extension of the cascade. However, we have verified that, for all of the tested IGMF strengths and time delays (up to one year), the cascade signal is well within the point spread function (PSF) of the \Fermilat (see right panel of Fig.~\ref{fig:cascade_and_map}), so that we can consider the cascade emission as point like. 

To quantify the impact of the EBL choice we also ran simulations using the models by \citet{Dominguez11}, \citet{Finke10}, and \citet{Saldana21}. Figure~\ref{fig:ebl_test} shows a comparison of the cascade SEDs obtained with each model, assuming a log-parabola VHE intrinsic spectrum without a cutoff and an IGMF strength of $\log_{10}(B/ \mathrm{G})=-20$. The differences found among models are minimal and do not affect the determination of the IGMF constraint. 

\section{\fermilat Analysis Details}
\label{app:LAT}

The analysis for the present work is done using Fermitools (version v2.2.0) and fermipy (version v1.2.2). We include LAT observations taken from $T_0$ + 9630 s to a year after the GRB, selecting P8R3 SOURCE events (evclass = 128, evtype = 3) with energies between 100 MeV and 1 TeV. We apply a maximum zenith angle cut at $90^{\circ}$ to prevent Earth limb contamination, and employ 5 bins in azimuth to reduce its impact on the exposure for short timescales \cite{Meyer19}. To adequately consider the detected emission at GeV energies, we bin the data following Table 3 in \cite{LATGRB221009A}. We complete it with a single bin from $T_0+3.3$ d to one year.

Each time-bin is independently analyzed with a binned maximum-likelihood framework \cite{Mattox96}. The Region of Interest (RoI) is defined as a $10^{\circ}\times10^{\circ}$ square centered at the GRB's nominal position (RA=288.21, DEC=19.73; \cite{Bissaldi22}), with $0.1^{\circ}$-width spatial bins and 8 bins per logarithmic energy decade. As our background, we include the standard Galactic and isotropic diffuse components (\texttt{gll\_iem\_v07} and \texttt{iso\_P8R3\_SOURCE\_V3\_v1}, respectively), and a component for each source in the latest catalog (4FGL-DR4 v34; \cite{4FGL, 4FGL-DR4}) within a region of $15^{\circ}\times15^{\circ}$. We employ the instrument response function P8R3\_SOURCE\_V3, although we note that energy dispersion is not applied to the isotropic component. The significance of each source is assessed through the test statistic TS $= -2 \ln(L_0/L_1)$, where $L_0$ is the log-likelihood of the null hypothesis and $L_1$ the log-likelihood of the complete model. 

The background is fitted by (1) optimizing the RoI by means of a standard iterative process\footnote{\url{https://fermipy.readthedocs.io/en/latest/fitting.html\#id2}}, (2) removing those components with a preliminary detection TS $<4$ or a predicted number of counts lower than 3, and (3) re-fitting the normalization of all sources within $3^{\circ}$ of GRB~221009A. For the time bin from $T_0+3.3$ d to one year, the final likelihood (for each IGMF strength) is obtained by including an additional pointlike source whose spectrum is described with results from the CRPropa cascade simulations. For the other time bins we consider three distinct scenarios:

\begin{enumerate}
    \item \textit{Cascade only}: A single source with the simulated cascade spectrum is added to the RoI. The spectrum is fixed to that obtained in the simulations (both normalization and shape). 
    \item \textit{Phenomenological afterglow with cascade}: In addition to the simulated cascade, an additional component for a pointlike source is added to account for the GeV emission from the GRB. Its spectrum is described with a power law with $\Gamma=2$ \cite{LATGRB221009A} and, in contrast to the cascade component, its normalization is left free in the fit.
    \item \textit{Physical afterglow with cascade}: In addition to the simulated cascade, an additional component for a pointlike source is added to account for the GeV emission, whose normalization and spectral shape is also fixed to that predicted by the afterglow model during the corresponding time interval. 
\end{enumerate}

\begin{figure*}
    \centering
    \includegraphics[width=0.49\linewidth]{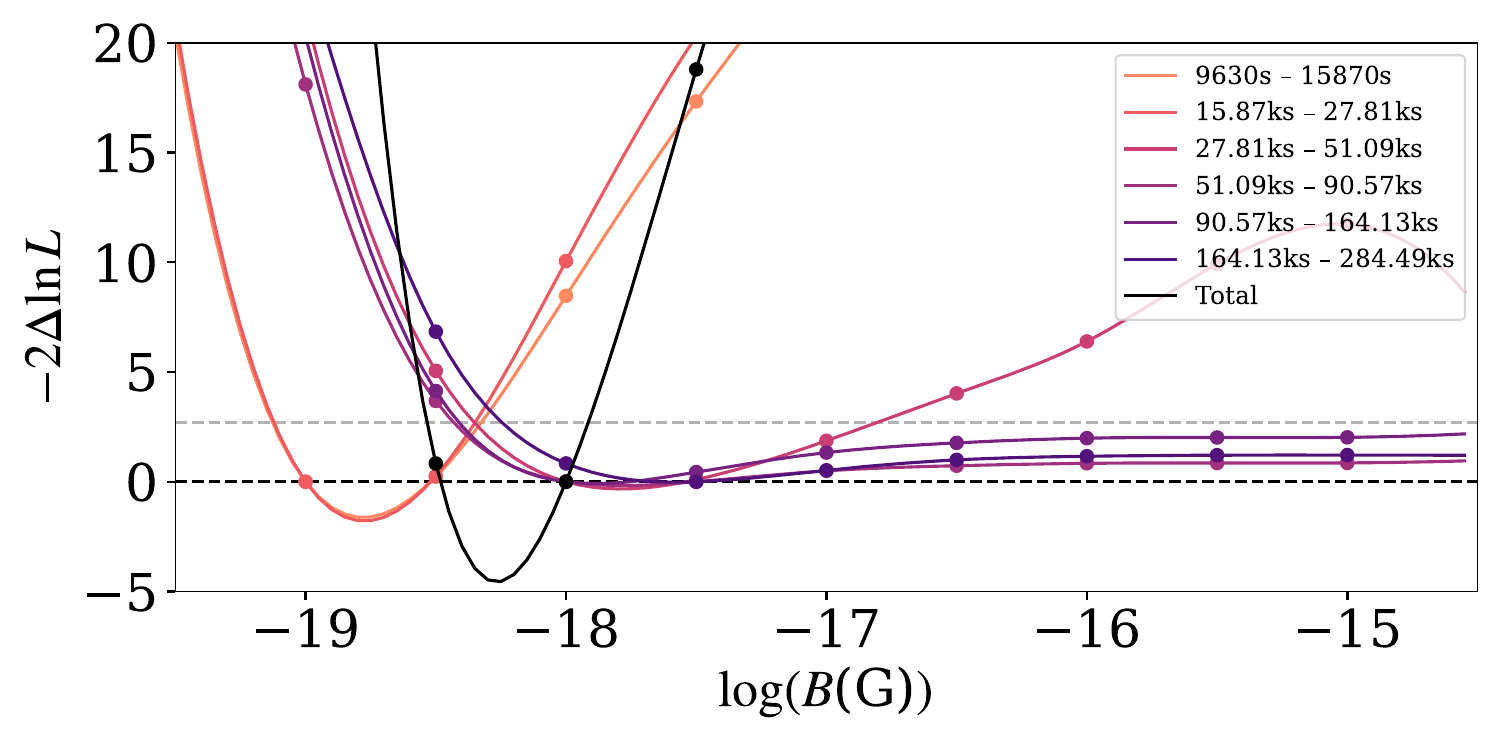}
    \includegraphics[width=0.49\linewidth]{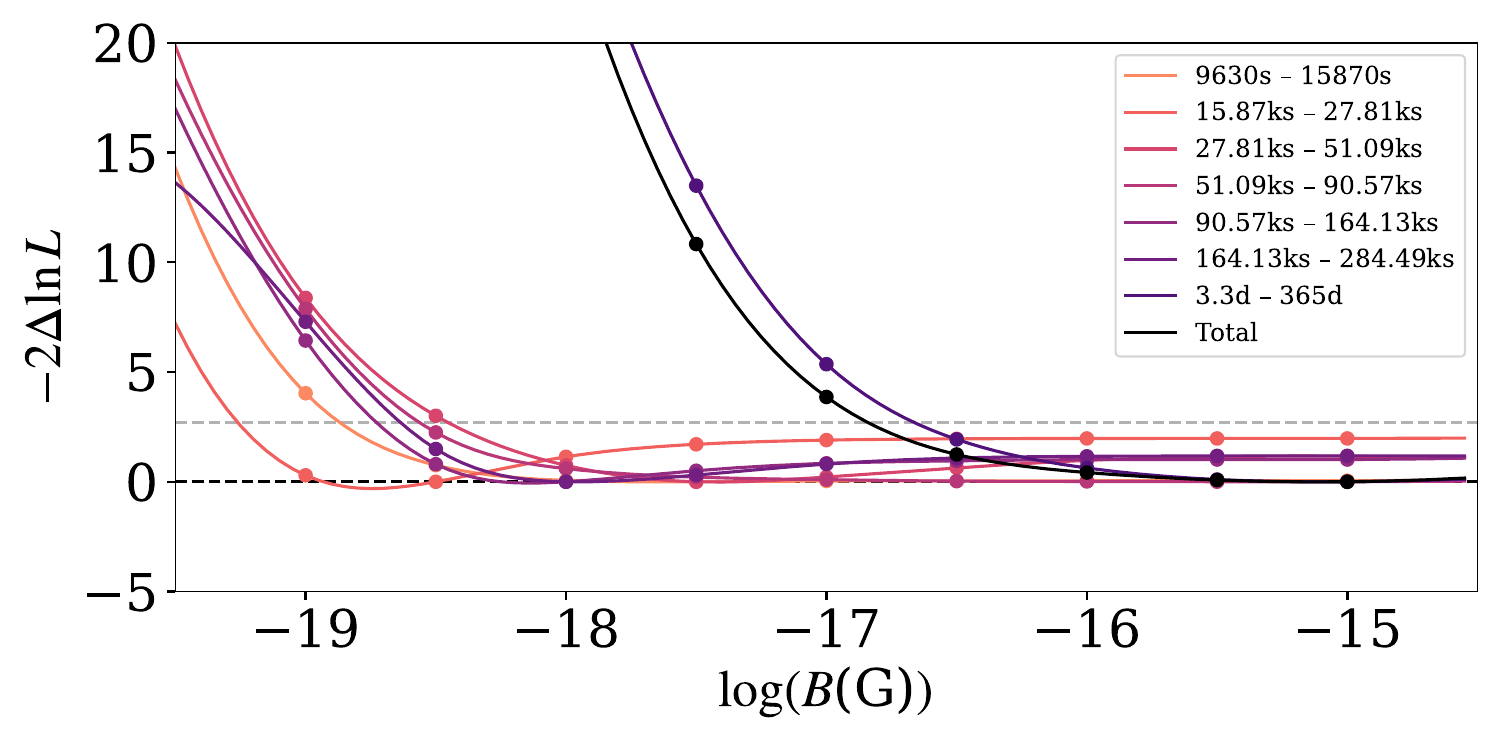}
    \includegraphics[width=0.49\linewidth]{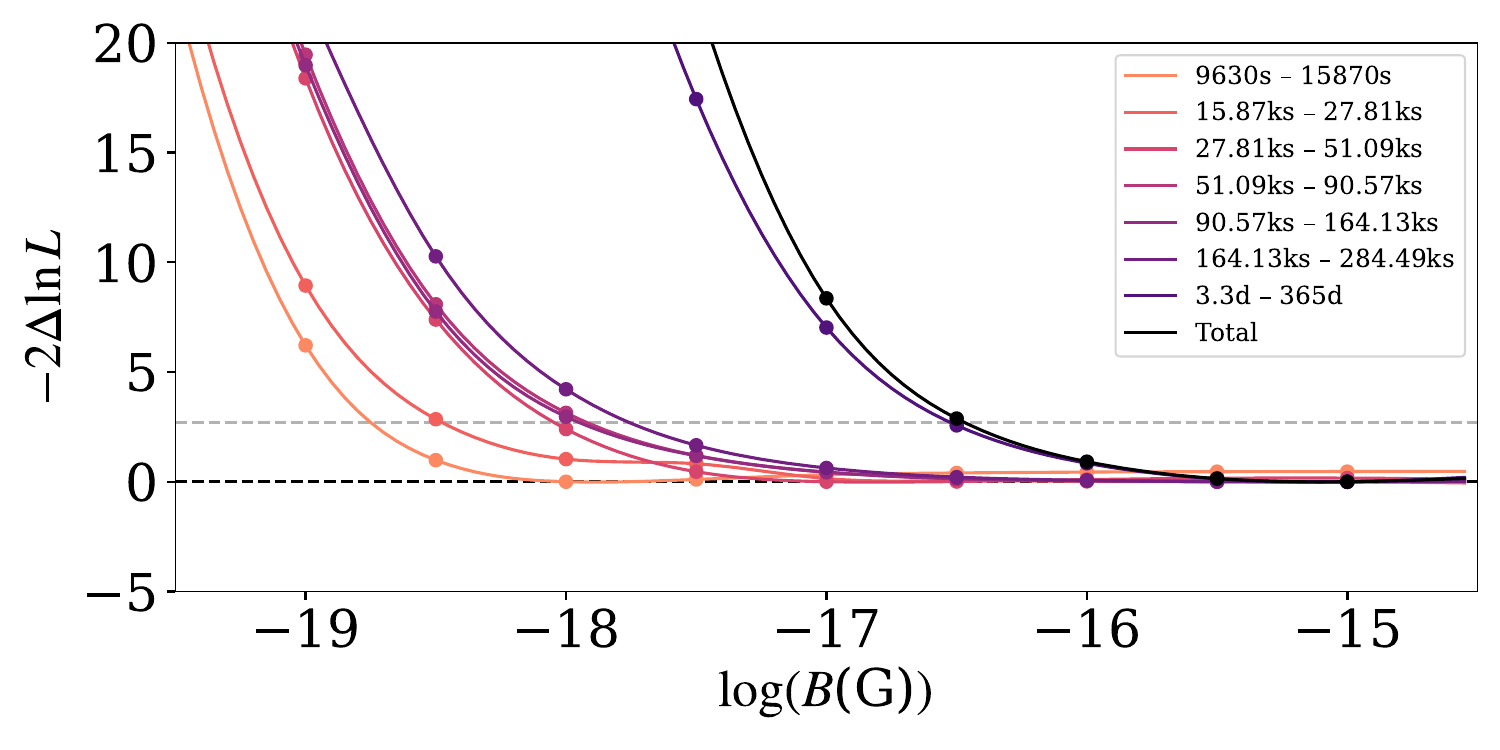}
     \includegraphics[width=0.49\linewidth]{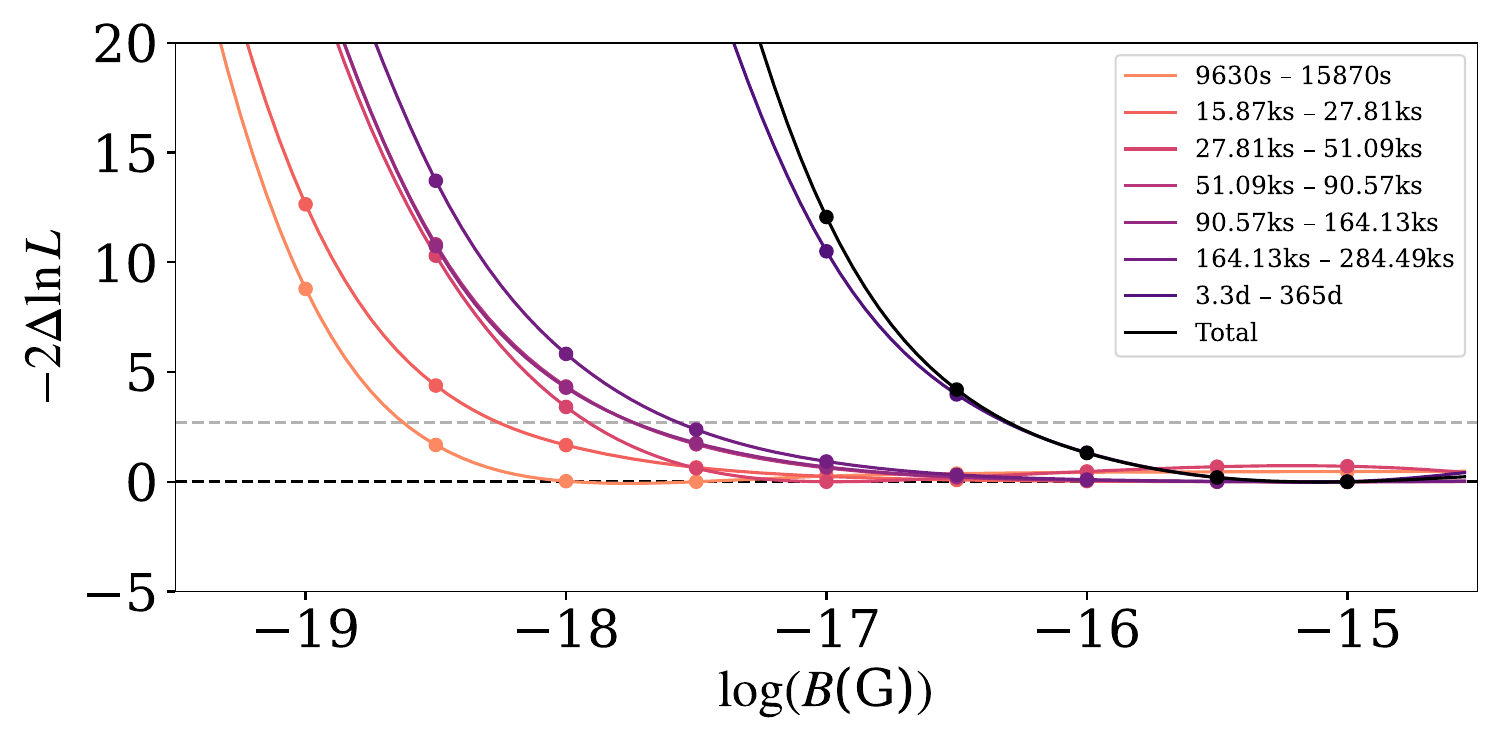}
    \caption{Likelihood profiles for further cases. Top panels display the likelihood profiles obtained by not extracting any signal (left) and by subtracting an afterglow component according to the physical model in Appendix \ref{app:SSCmodel} (right). Both cases assume an exponential cutoff at 7 TeV. Bottom panels display the profiles derived assuming an \textit{ad hoc} afterglow emission with $\Gamma=2$, but with the cutoff at 13 TeV (left) or without it (right). Solid lines represent cubic spline interpolations between the derived likelihood values.}
    \label{fig:likelihoods_other cases}
\end{figure*}

The likelihood profiles for different scenarios are presented in Fig.~\ref{fig:likelihoods_other cases}, with the complete list of limits derived for each scenario summarized in Tab.~\ref{tab:IGMF_lower_bounds_7TeV}. Additionally, we present the SEDs for a pointlike source at the position of GRB 221009A with and without a power law component extracting the GeV signal (Fig.~\ref{fig:cascade_seds_with_afterglow}). Fluxes for the afterglow component are fully consistent with those derived in \cite{LATGRB221009A}.

%\begin{sidewaystable}[ht]
\begin{table*}[ht]
\centering
%\begin{tabular}{ |p{3.5cm}||p{2cm}|p{2cm}| }
\begin{tabular}{ r|| p{2.2cm}|p{2.2cm} | p{2.2cm}|p{2.2cm} | p{2.2cm}|p{2.2cm} }
 %\hline
 %\multicolumn{4}{|c|}{Country List} \\
 \hline
 \hline
  \multirow{2}{*}{$T_0$-$T_f$}  & \multicolumn{2}{c}{cutoff at 7~TeV} & \multicolumn{2}{c}{cutoff at 13~TeV} & \multicolumn{2}{c}{No cutoff}\\
  & power law & GRB model & power law & GRB model & power law & GRB model \\
 \hline
 9630 s -15870 s  & $1.4 \times 10^{-19}$ & $1.4 \times 10^{-19}$  & $1.8 \times 10^{-19}$ & $1.8 \times 10^{-19}$  & $2.2 \times 10^{-19}$ & $2.5 \times 10^{-19}$ \\
 15870 s - 27810 s  &  $2.5 \times 10^{-19}$ & $5.6 \times 10^{-20}$ &  $3.2 \times 10^{-19}$ & $7.1 \times 10^{-20}$  &  $5.6 \times 10^{-19}$ & $1.1 \times 10^{-19}$ \\
 27810 s - 51090 s & $7.1 \times 10^{-19}$ & $3.5 \times 10^{-19}$ & $9.0 \times 10^{-19}$ & $5.6 \times 10^{-19}$ & $1.3 \times 10^{-18}$ & $7.9 \times 10^{-19}$\\
 51090 s - 90570  s & $9.0 \times 10^{-19}$ & $2.8 \times 10^{-19}$ & $1.3 \times 10^{-18}$ & $4.0 \times 10^{-19}$ & $1.8 \times 10^{-18}$ & $7.1 \times 10^{-19}$\\
 90570 s - 164130 s & $7.9 \times 10^{-19}$ & $1.8 \times 10^{-19}$ & $1.1 \times 10^{-18}$ & $2.5 \times 10^{-19}$ & $1.8 \times 10^{-18}$ & $4.0\times 10^{-19}$ \\
 164130 s - 284490 s & $1.3 \times 10^{-18}$ & $2.2 \times 10^{-19}$ & $1.8 \times 10^{-18}$ & $3.2 \times 10^{-19}$ & $2.8 \times 10^{-18}$ & $5.0 \times 10^{-19}$ \\
 3.3 days - 365 days & $2.2 \times 10^{-17}$ &  $2.2 \times 10^{-17}$ & $3.2 \times 10^{-17}$ & $3.2 \times 10^{-17}$  & $5.0 \times 10^{-17}$ & $5.0 \times 10^{-17}$\\
 \hline\hline
 combined & $2.5 \times 10^{-17}$ & $1.4 \times 10^{-17}$ & $3.5 \times 10^{-17}$ & $1.8 \times 10^{-17}$ & $5.0 \times 10^{-17}$ & $2.5 \times 10^{-17}$ \\
 \hline
\end{tabular}
\caption{The lower bounds on the IGMF strength, expressed in G at the 95\% confidence level, are presented for all the tested time intervals. These results assume the power law and GRB physically motivated models as spectral models for the GRB in the \fermilat analysis. The results were obtained using the log-parabola model, Eq.~\eqref{eqn:logpar}. Exponential cutoffs are considered by including an additional term to the equation.}
\label{tab:IGMF_lower_bounds_7TeV}
\end{table*}
%\end{sidewaystable}

To estimate if similar studies would further improve the constraints by increasing the exposure time, we extrapolate the derived exposure map of the ROI up to 5 years. A $2\sigma$ confidence level sensitivity curve is derived, requiring at least 3 counts from the echo emission and considering the impact of diffuse components. We derive such sensitivity curves for narrow energy bands, with four bins per energy decade and assuming a flat spectrum with $\Gamma=2$ within each energy bin. We note that these narrow bands are required to account for the energy-dependent time evolution of the cascade. Additional CRPropa simulations up to $T_0+5$ yr predict a measurable (yet moderate) improvement in the sensitivity from 3 to 15 GeV (Fig. \ref{fig:Sensitivity}), the most favorable energy range for the \fermilat.

\begin{figure*}[h!]
    \centering
    \includegraphics[width=0.46\linewidth]{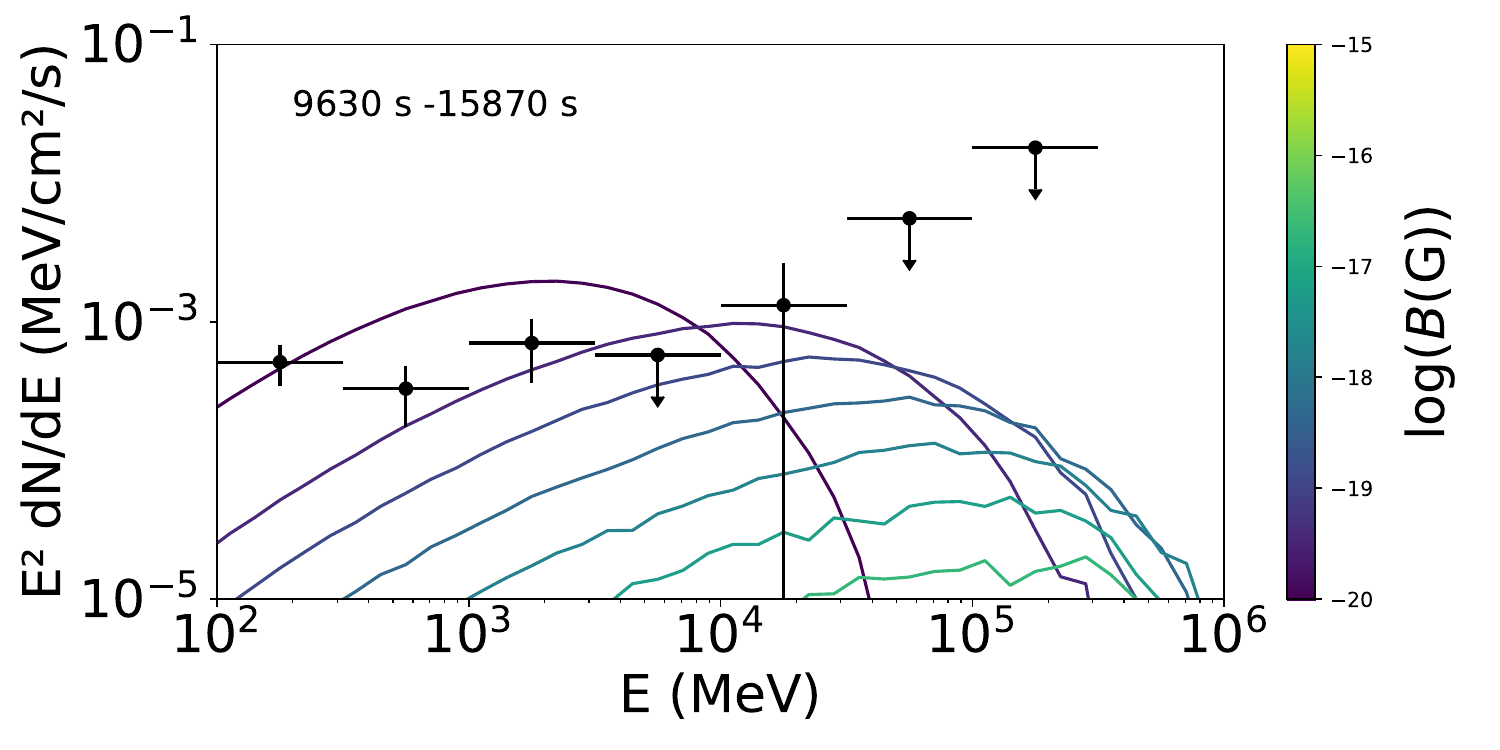}
    \includegraphics[width=0.45\linewidth]{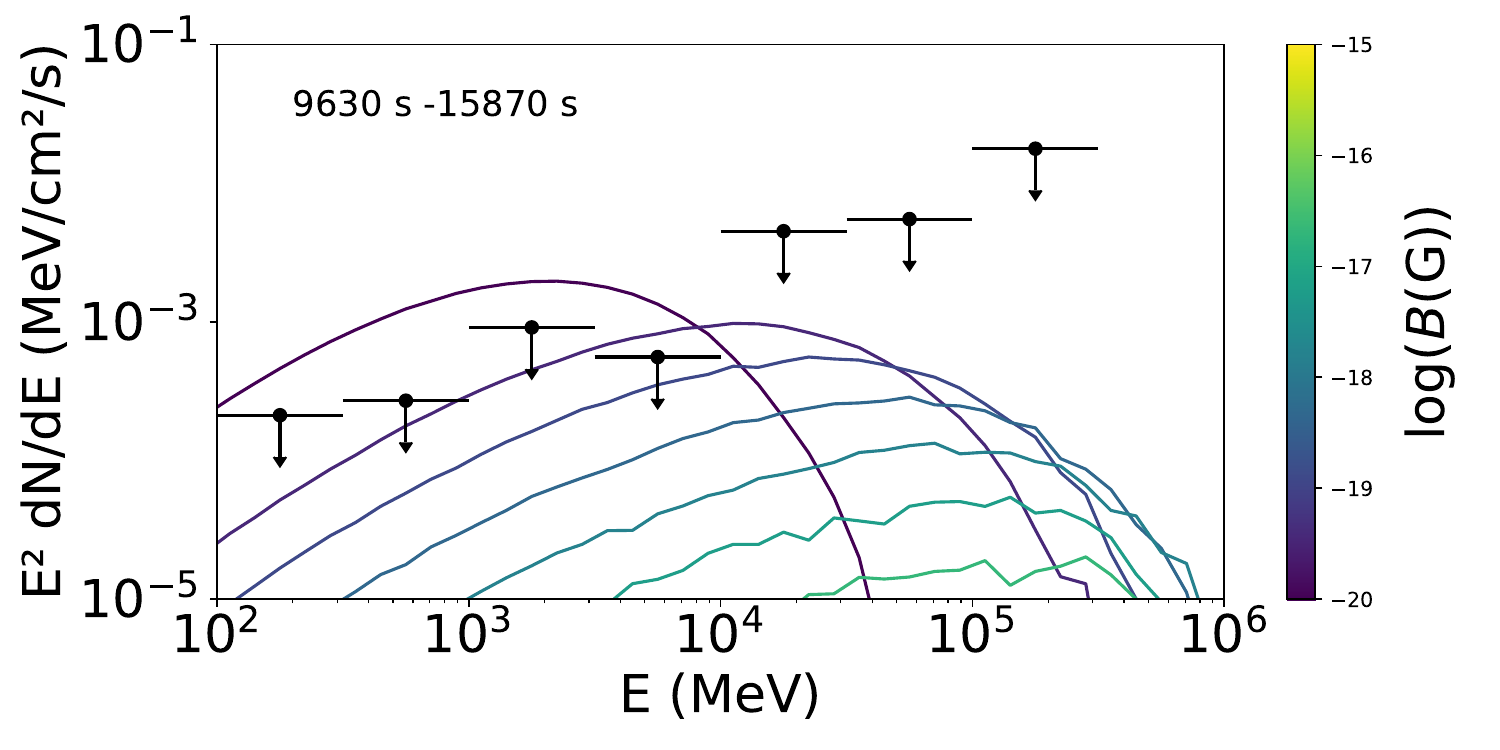}
    \includegraphics[width=0.46\linewidth]{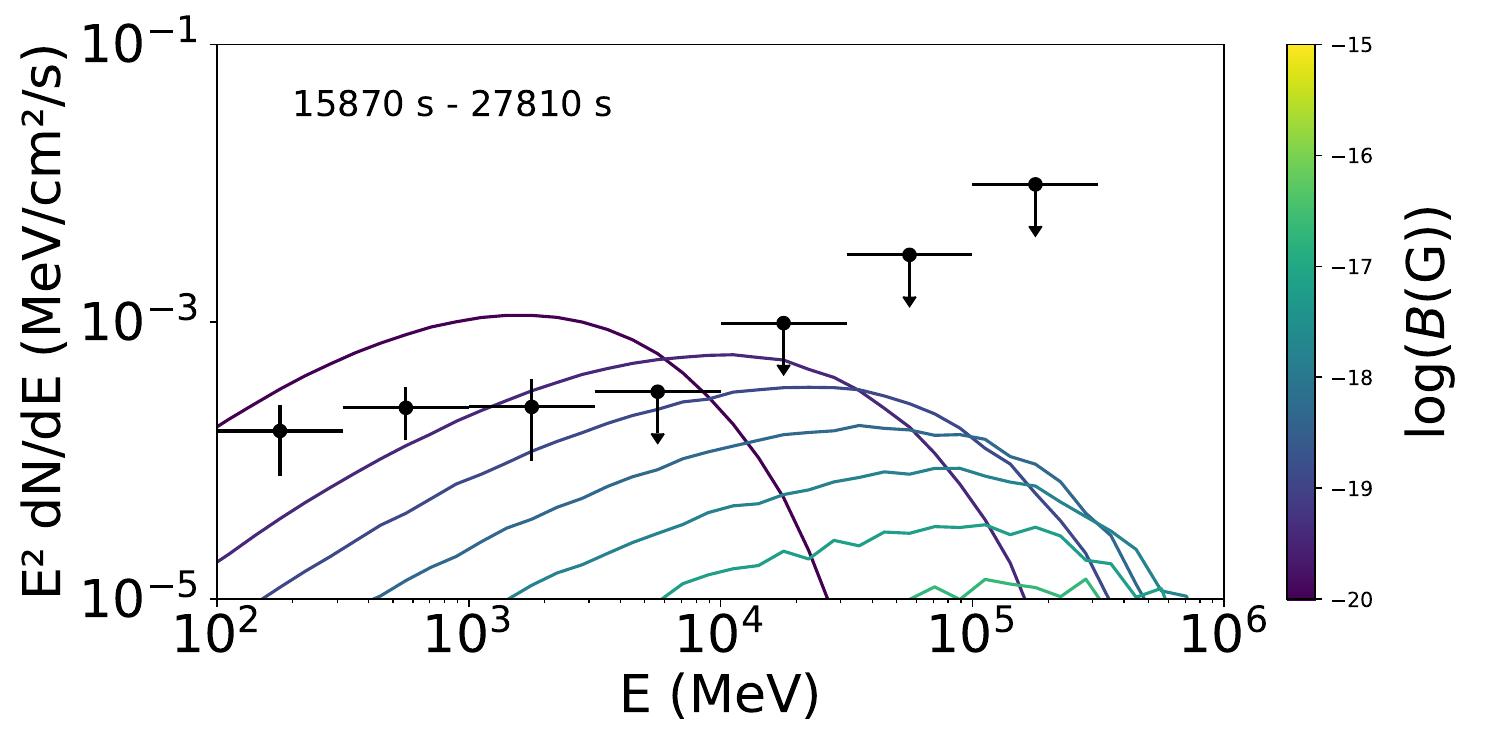}
    \includegraphics[width=0.45\linewidth]{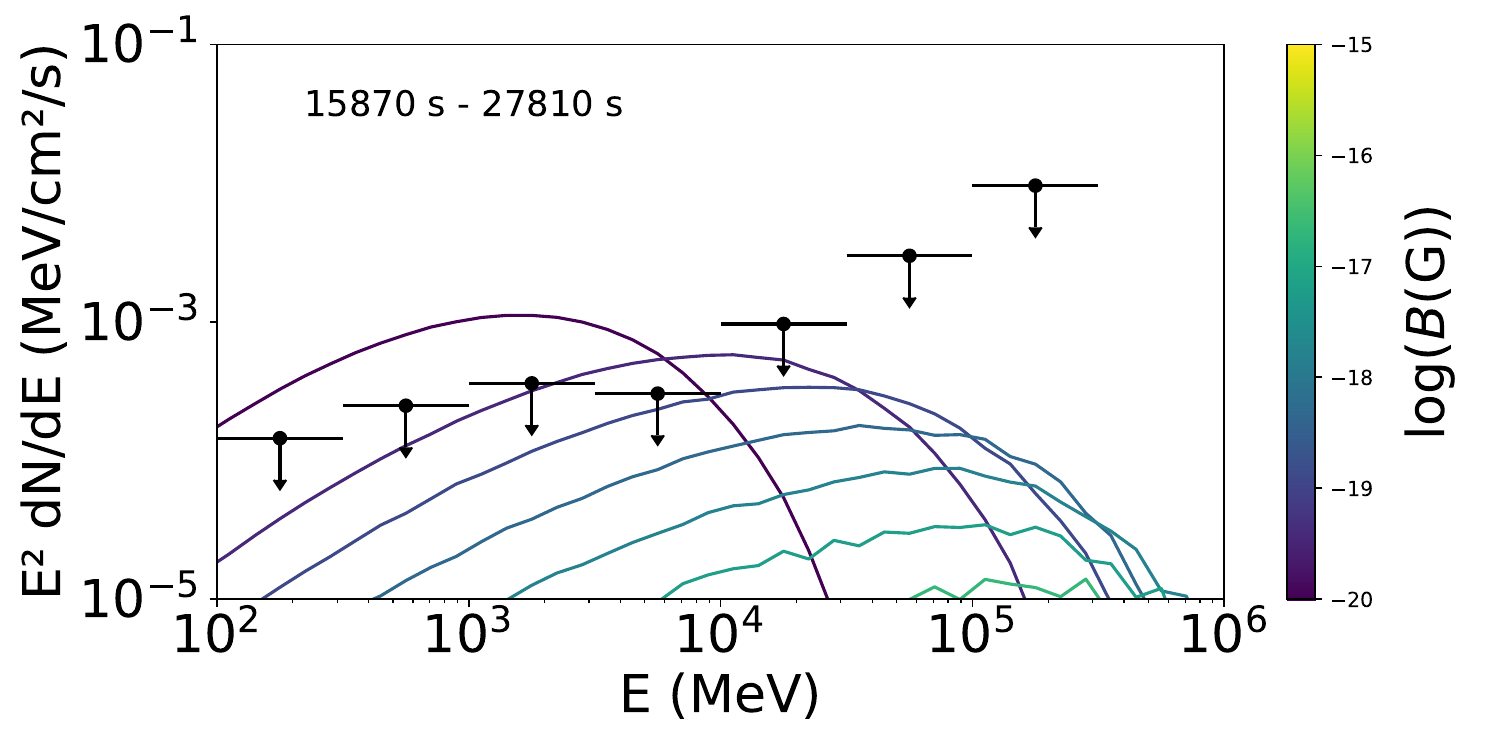}
    \includegraphics[width=0.46\linewidth]{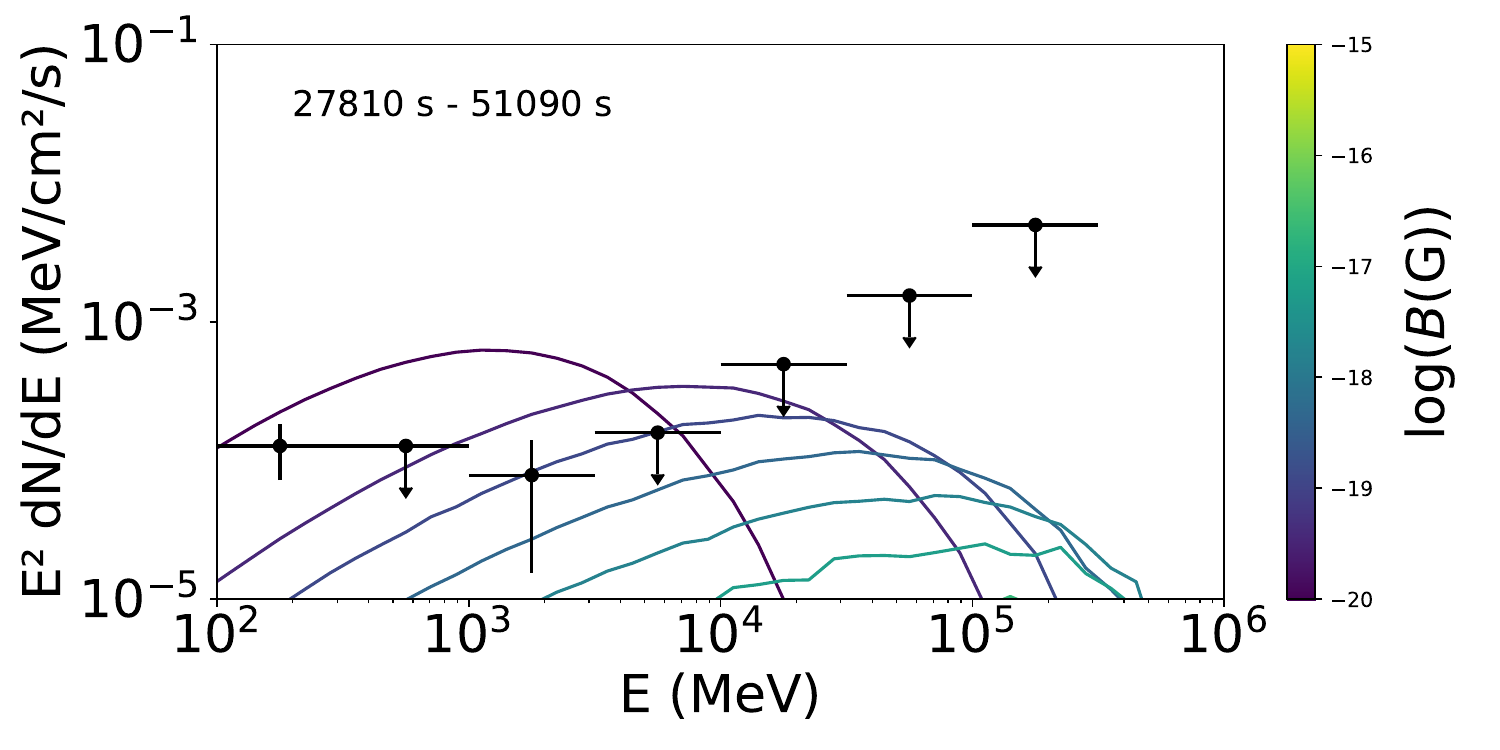}
    \includegraphics[width=0.45\linewidth]{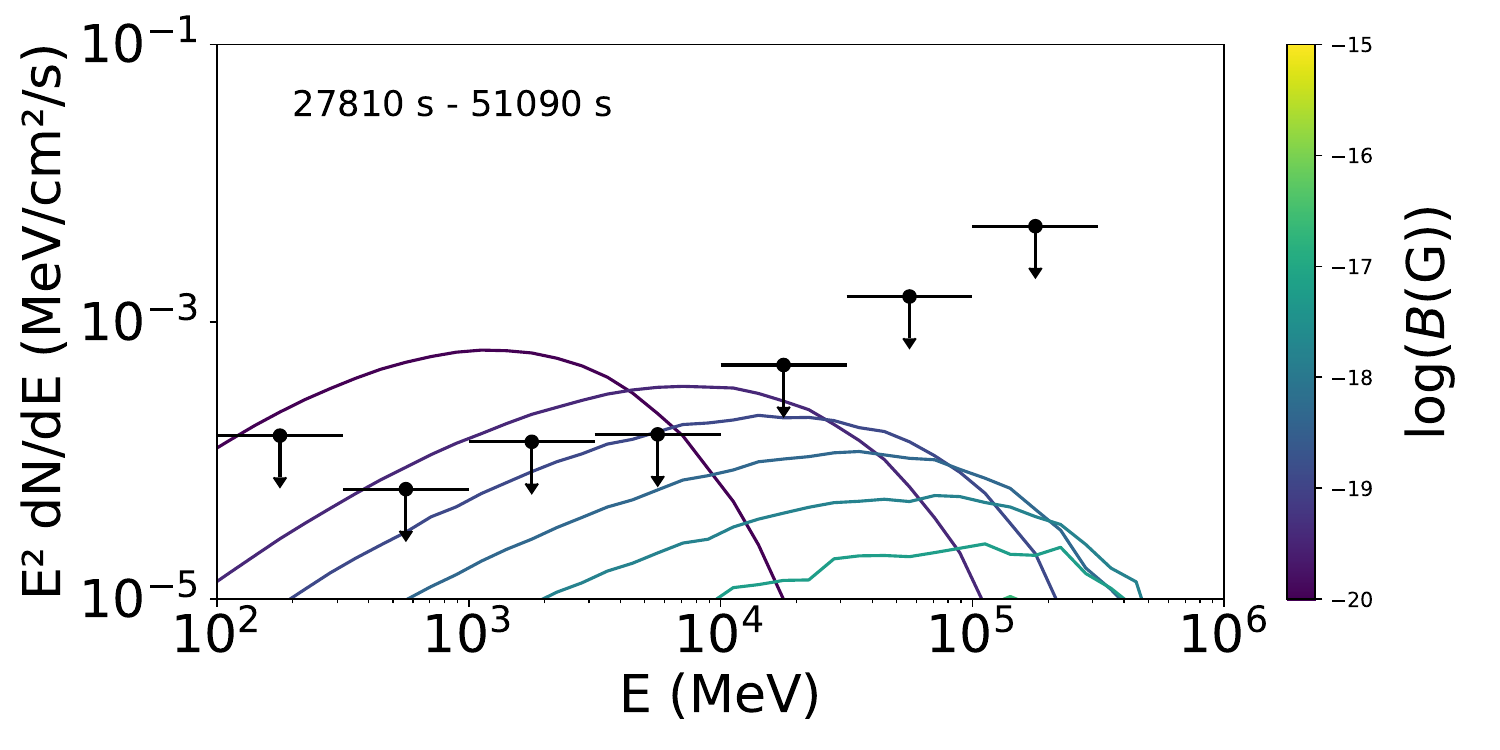}
    \includegraphics[width=0.46\linewidth]{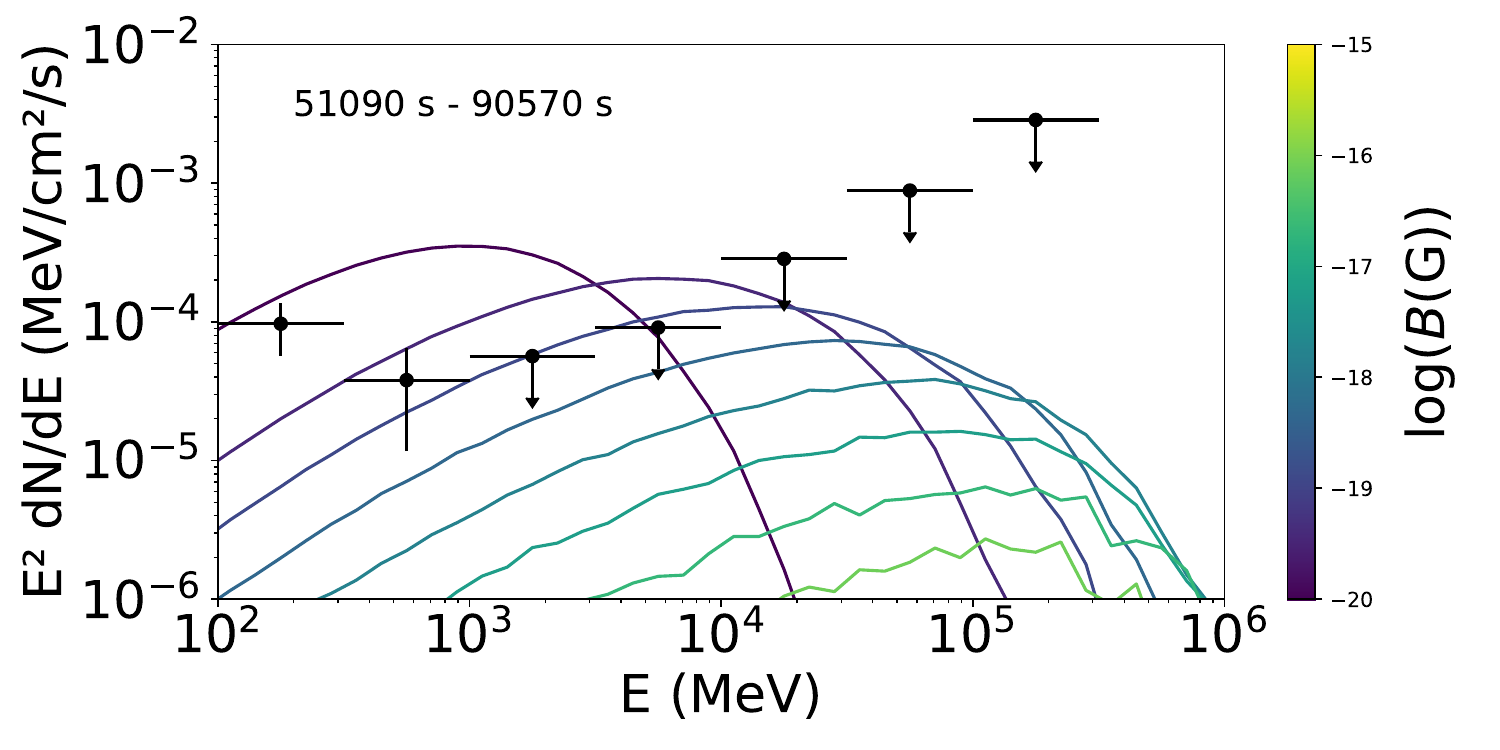}
    \includegraphics[width=0.45\linewidth]{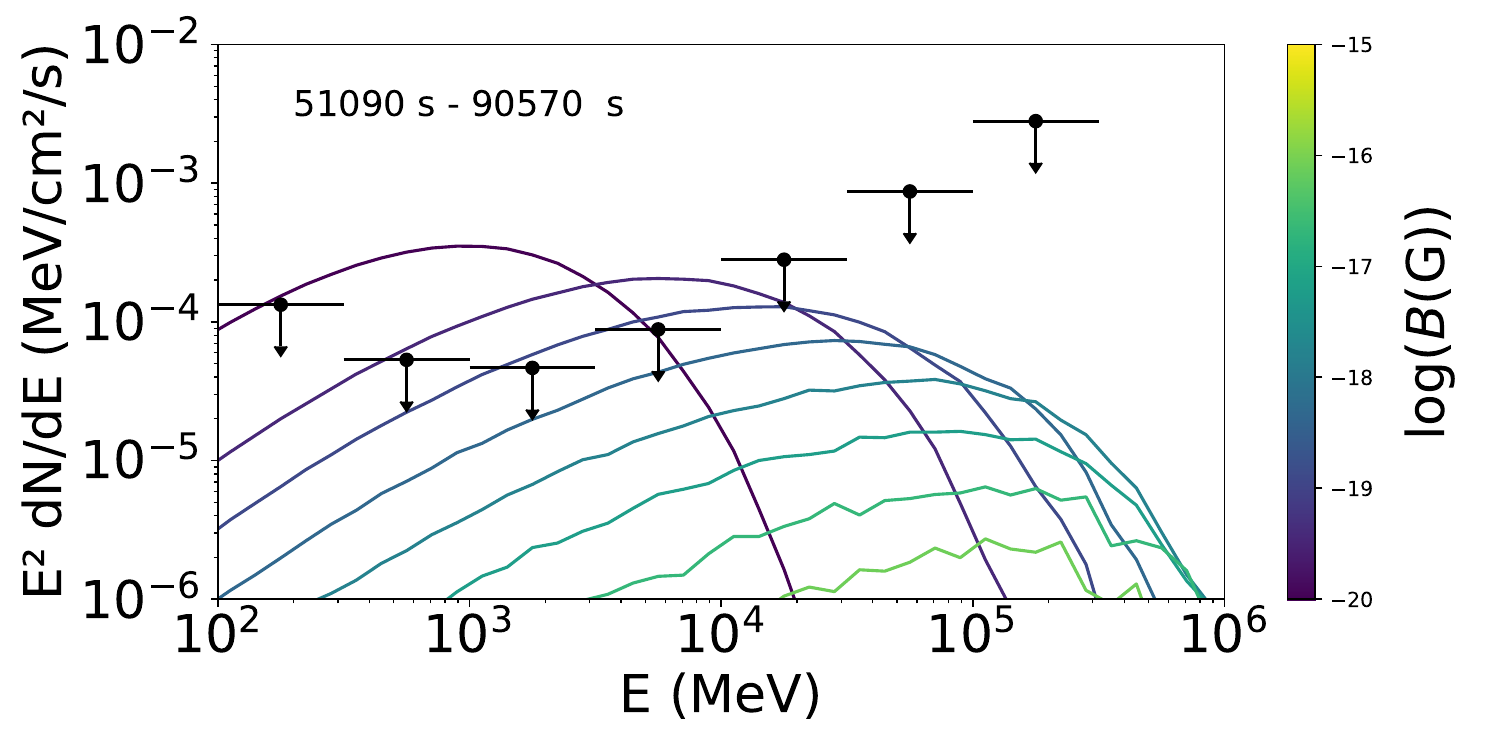}
    \includegraphics[width=0.46\linewidth]{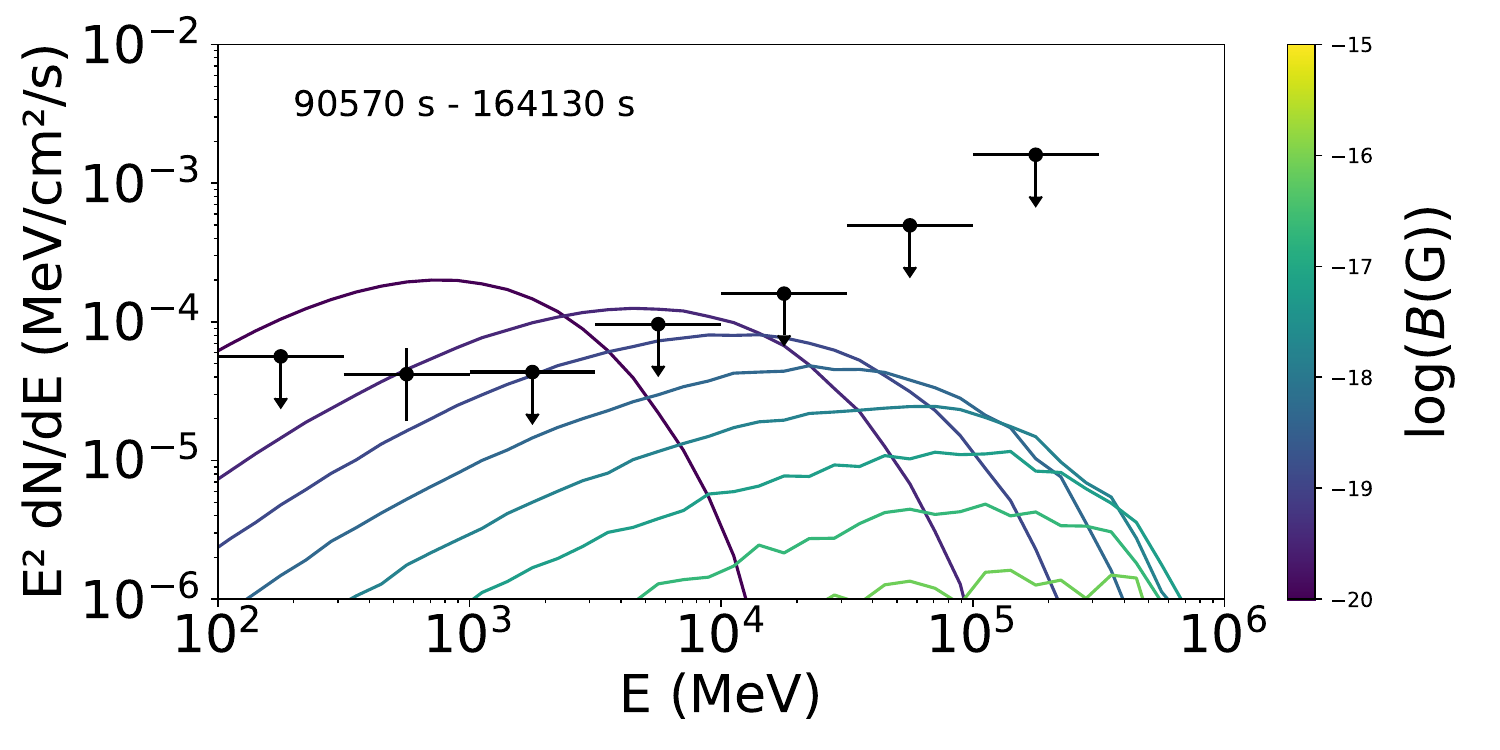}
    \includegraphics[width=0.45\linewidth]{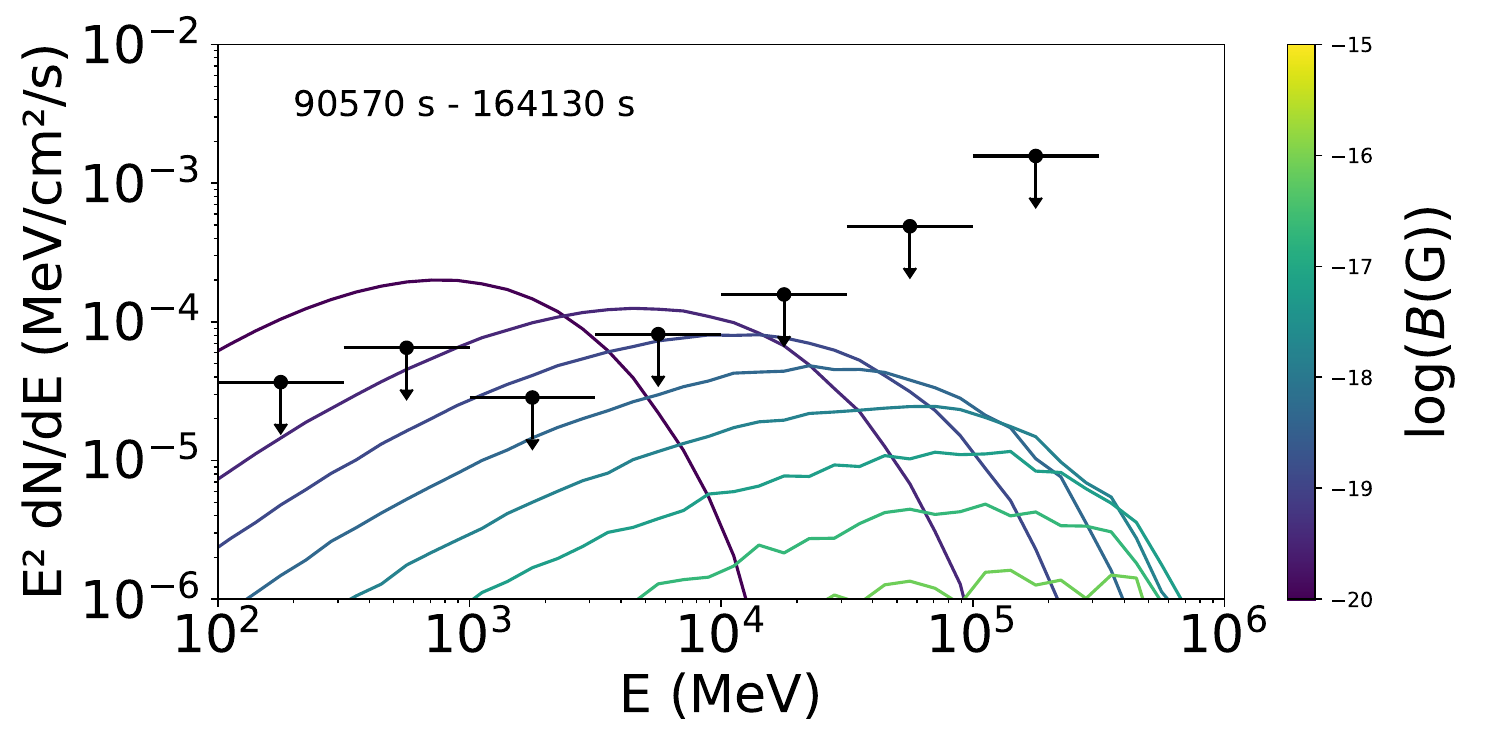}
     \caption{\textit{Left}: SED cascades for the different time bins considered, in comparison with the LAT observations during the corresponding time window. \textit{Right}: Same SED cascades in comparison with the upper limits derived from LAT observations once the afterglow emission was subtracted with a power law component during the corresponding time window.}\label{fig:cascade_seds_with_afterglow}
\end{figure*}

\begin{figure*}[h!]
    \centering
    \ContinuedFloat
    \includegraphics[width=0.46\linewidth]{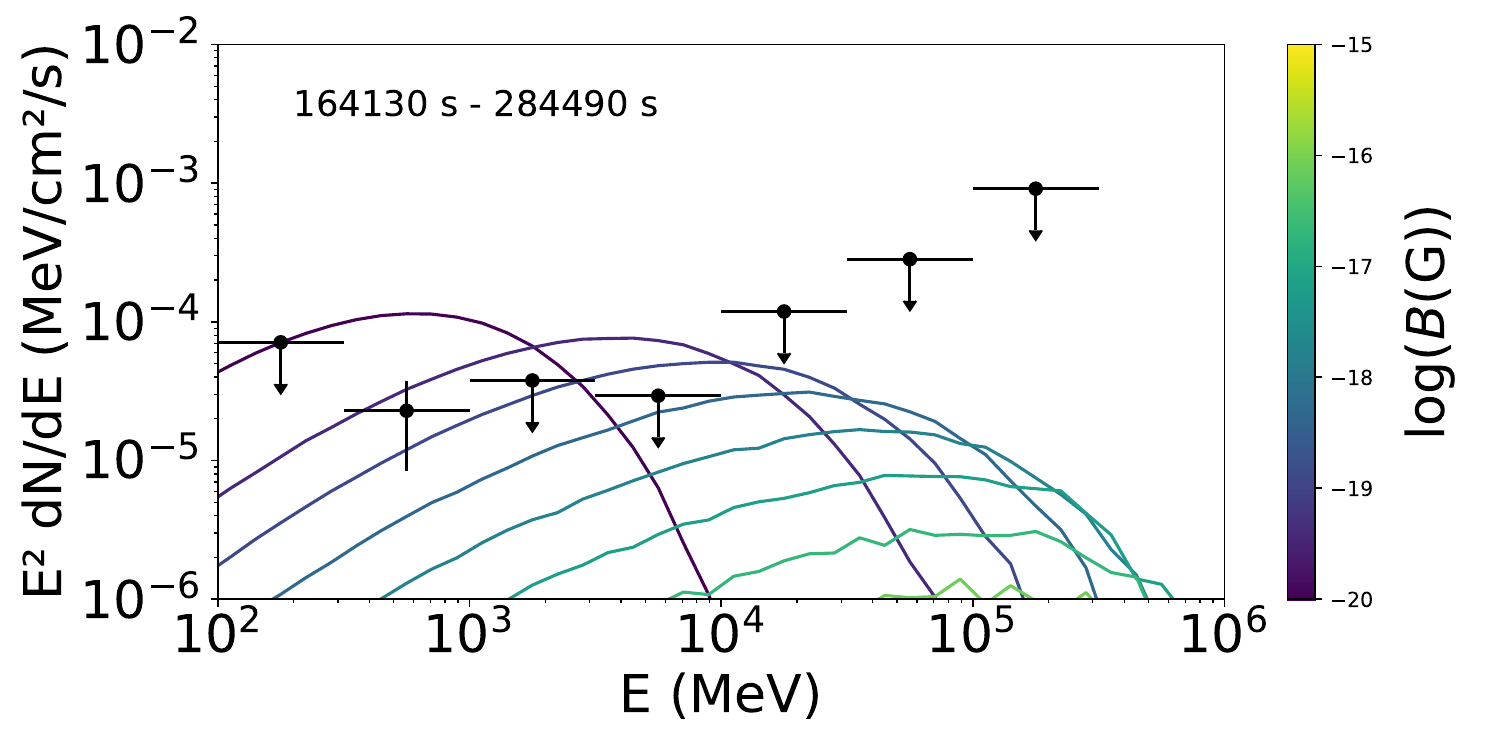}
    \includegraphics[width=0.45\linewidth]{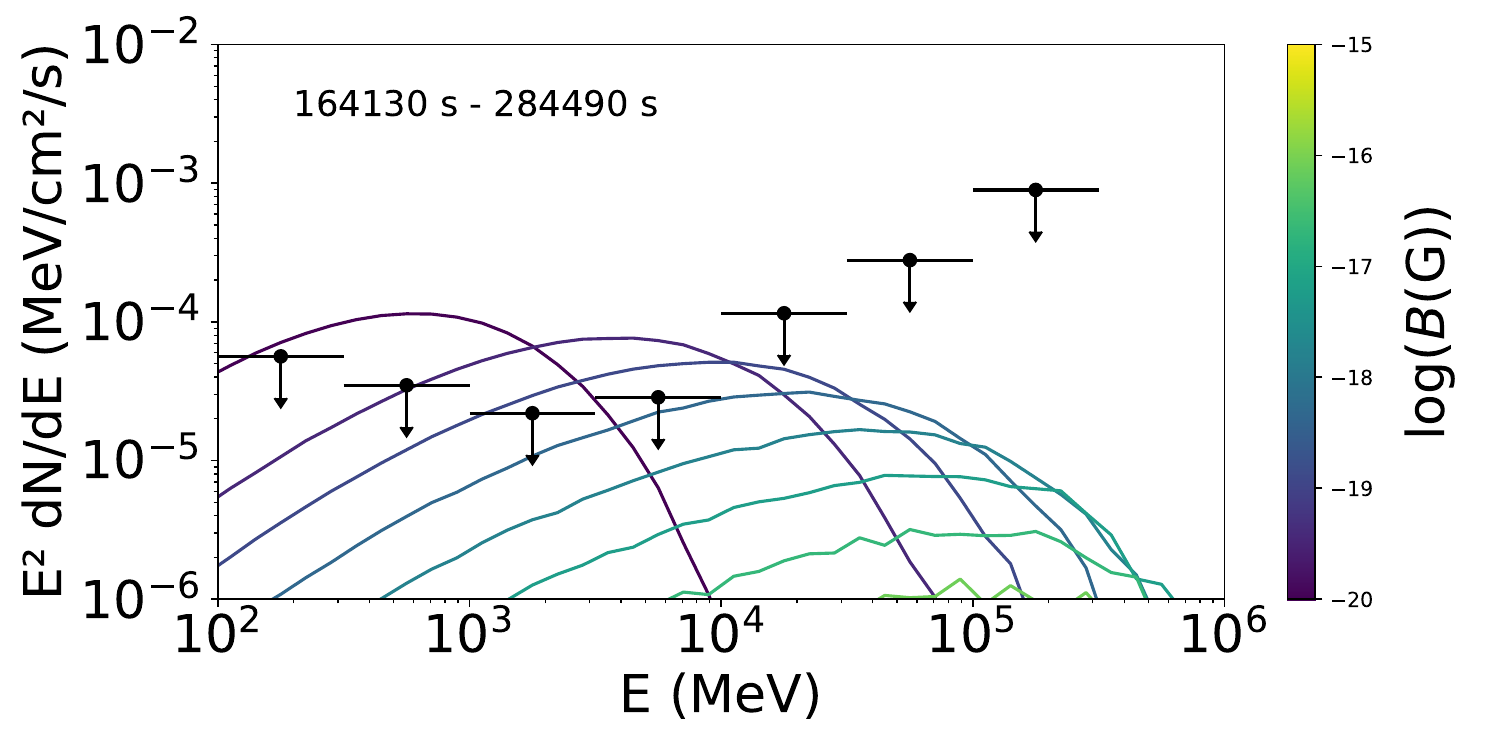}
    \caption{(Continued)}\label{fig:cascade_seds_with_afterglow}
\end{figure*}

\begin{figure*}
    \centering
    \includegraphics[width=0.49\linewidth]{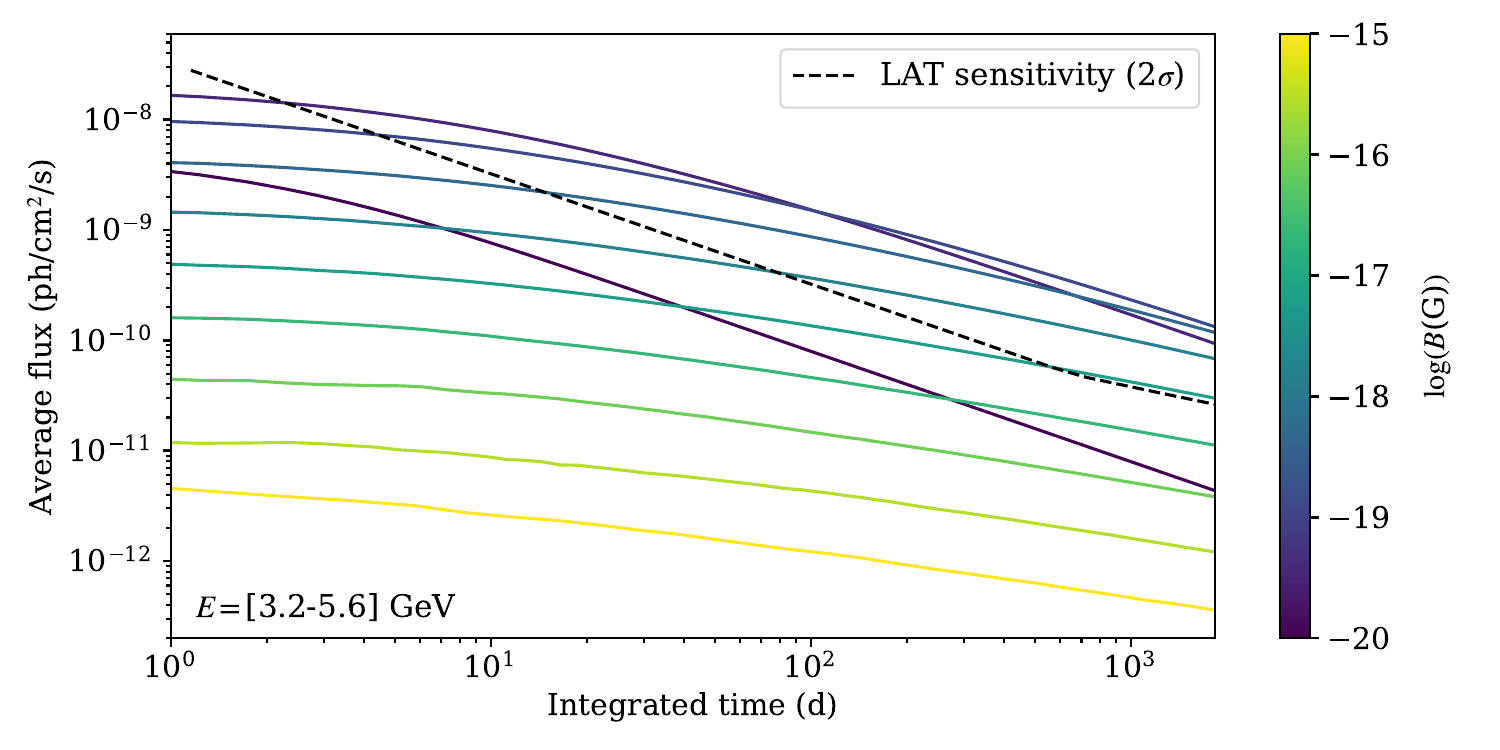}
    \includegraphics[width=0.49\linewidth]{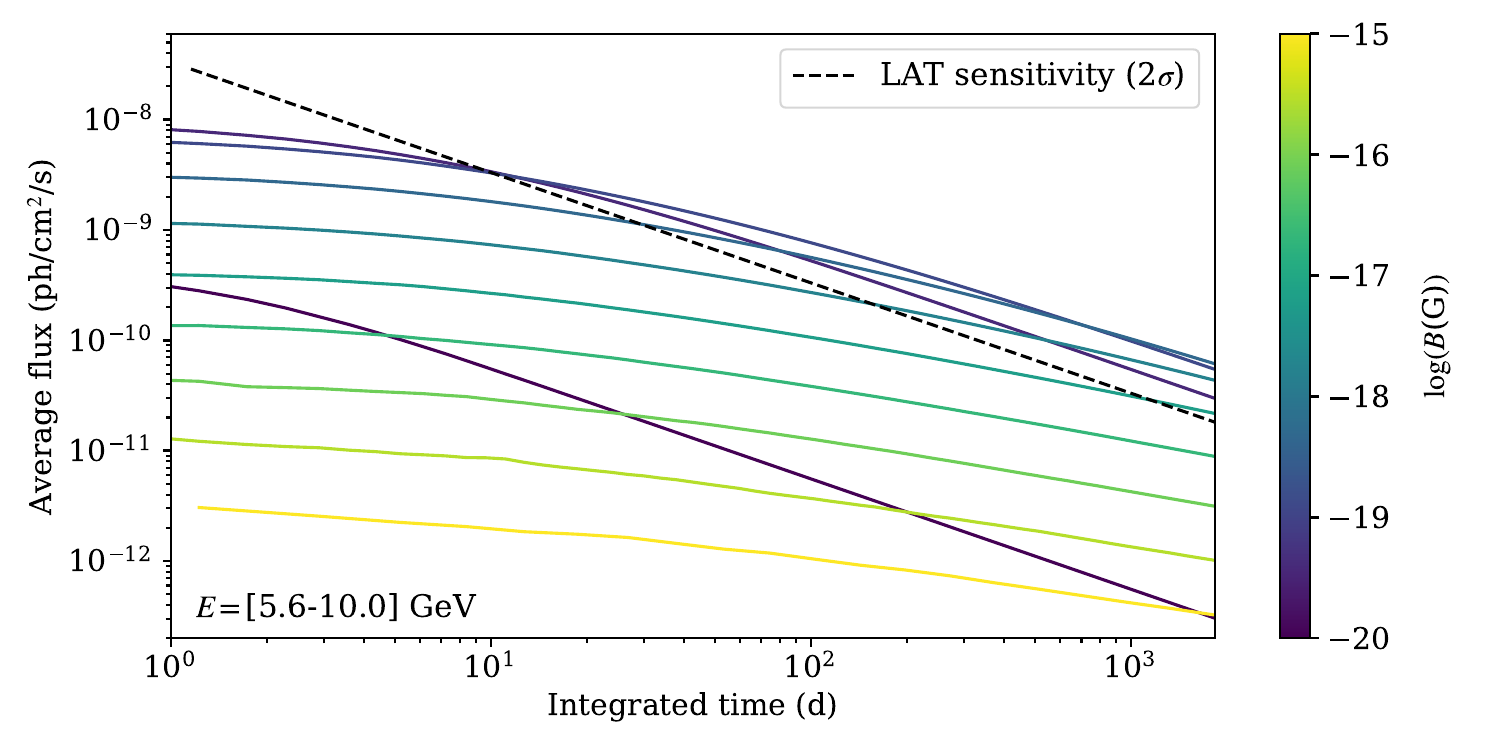}
    \includegraphics[width=0.49\linewidth]{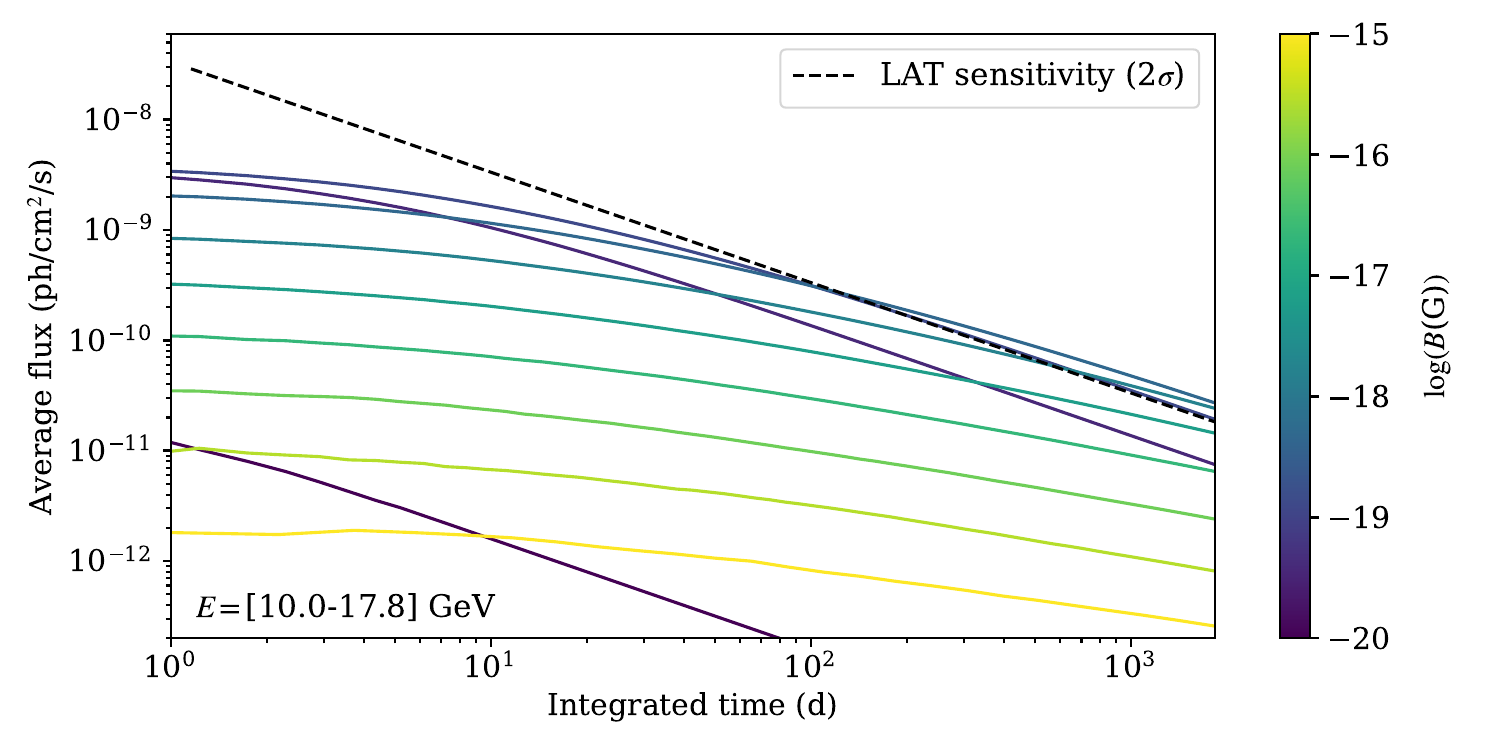}
    \caption{\fermilat sensitivity (95\% confidence level) as a function of observation time compared with the average photon flux expected from the echo emission for the same integration times. The results are shown for three narrow energy bands only.
    }
    \label{fig:Sensitivity}
\end{figure*}

\section{SSC afterglow model at GeV~energies} \label{app:SSCmodel}
We model the afterglow as leptonic synchrotron and synchrotron self-Compton (SSC) emission from the forward shock. Semi-analytical approximations for SSC were calculated by \cite{Sari+01ic,panaitescu00}. Here we follow the treatment of \cite{chiangdermer99} and numerically calculate the spectral energy distribution at the time of the LHAASO and \textit{Fermi}-LAT observations. 

We numerically solve for the bulk Lorentz factor of the blast wave, $\Gamma$, as a function of radius and follow the distribution of the electrons' random Lorentz factor, $\gamma_e$ through the continuity equation. We assume the electrons develop a power law distribution as they are injected into the blast wave from the interstellar medium. We take into account radiative and adiabatic cooling. 

\begin{figure*}
    \centering
    \includegraphics[width=0.8\linewidth]{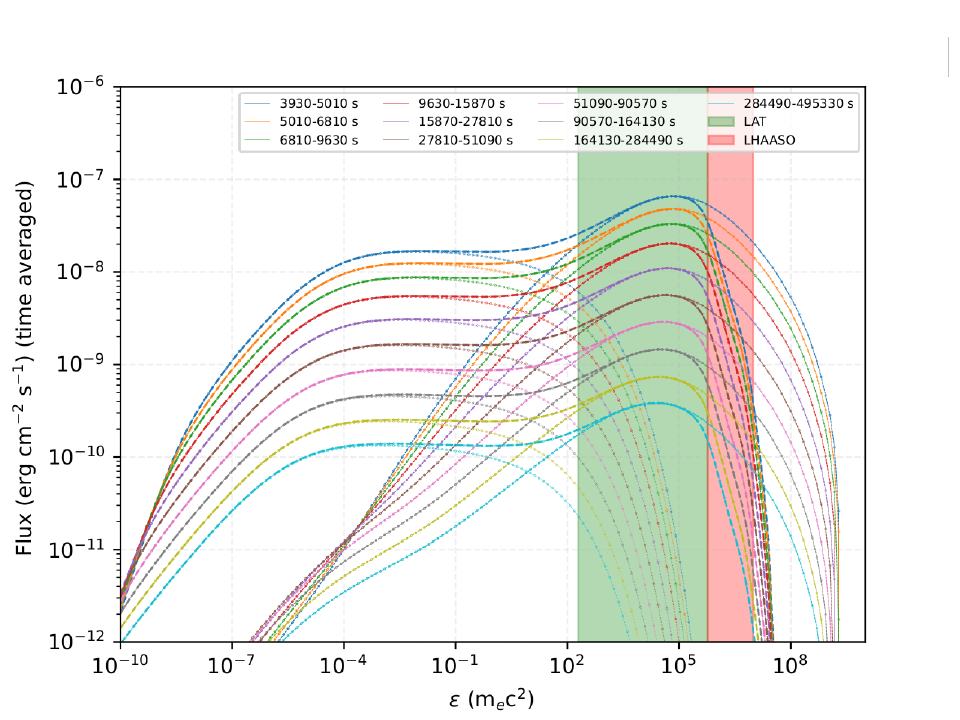}
    \caption{SEDs for the SSC model during the time bins defined in Table 3 of \cite{LATGRB221009A}. Each color represents a different time bin. The two components correspond to synchrotron and SSC emission (dotted and dash-dotted lines, respectively). Dashed lines represent the total emission once EBL absorption is considered. The LAT energy range, in green, is used in our likelihood analysis for the \textit{physical afterglow with cascade} scenario.
    }
    \label{fig:SSC_SED_prediction}
\end{figure*}

To obtain the spectral energy distribution, at each time we convolve the synchrotron and SSC emissivity with the electron distribution. To compare the model with the LHAASO measurements, we integrate the flux over their reported time range. We vary the free parameters of the model (total energy $E_0$, initial Lorentz factor $\Gamma _0$, electron distribution index $s$, electron $\xi_e$ and magnetic field $\xi_B$ equipartition parameter) to match the observed LHAASO light curve (Tab. \ref{tab:SSC_model}). We do not perform a fit. Once we achieve an acceptable match to the LHAASO light curve, we integrate the model over the required time bin and derive the expected SED for the LAT energy range (Fig.~\ref{fig:SSC_SED_prediction}). 

\begin{table}
\centering
%\begin{tabular}{ |p{3.5cm}||p{2cm}|p{2cm}| }
\begin{tabular}{ r| p{3cm} }
 %\hline
 %\multicolumn{4}{|c|}{Country List} \\
 \hline
 \hline
 Parameter & Value \\
 \hline
 $E_0$ & $2.45 \times 10^ {55} $ erg \\
 $\Gamma _0$ & 530 \\
 $s$ & 2.05 \\
 $\xi_e$ & $2.5 \times 10^{-2}$ \\
 $\xi_B$ & $6.0 \times 10^{-4}$ \\
 $n_{\rm{ext}}$ & $0.4$ cm$^{-3}$ \\
 \hline
\end{tabular}
\caption{Parameters used in the leptonic model to account for the afterglow component at GeV energies.}
\label{tab:SSC_model}
\end{table}

\end{document}